\documentclass[a4paper,11pt]{article}

\usepackage[height=23cm, width=17cm, hmarginratio={1:1}]{geometry}

\usepackage{amsmath}
\usepackage{mathrsfs}
\usepackage{amsmath}
\usepackage{amssymb}
\usepackage{amsthm}
\usepackage{bm}
\usepackage{tikz-cd}
\usepackage{geometry}
\usepackage{appendix}
\usepackage{caption}
\usepackage{subcaption}
\usepackage{rotating}
\usepackage{wrapfig,framed,bbm}
\usepackage{graphicx,color}
\usepackage{hyperref}
\usepackage[numbers,sort&compress]{natbib}

\newcommand{\ad}{{\rm ad}}
\newcommand{\ke}{{\rm Ker\,}}
\newcommand{\im}{{\rm Im\,}}
\newcommand{\tr}{{\rm Tr\,}}

\newcommand{\ckp}{\textsc{\tiny\rm cKP}}

\newcommand{\p}{\partial}

\newcommand{\mat}{{\rm Mat}}

\newtheorem{theorem}{Theorem}[section]
\newtheorem{lemma}[theorem]{Lemma}

\newtheorem{remark}[theorem]{Remark}
\newtheorem{prop}[theorem]{Proposition}
\newtheorem{cor}[theorem]{Corollary}
\theoremstyle{example}
\newtheorem{example}[theorem]{Example}
\numberwithin{equation}{section}

\allowdisplaybreaks[1]

\def\be{\begin{equation}}
\def\ee{\end{equation}}

\begin{document}
	
\title{The constrained KP hierarchy and \\ the bigraded Toda hierarchy of $(M,1)$-type}
\author{Ang Fu, Di Yang, Dafeng Zuo}
\date{}
\maketitle

\begin{abstract}
In this paper, we extend the matrix-resolvent method to the study of the Dubrovin--Zhang type tau-functions for the constrained KP hierarchy  and the bigraded Toda hierarchy of $(M,1)$-type. 
We show that the Dubrovin--Zhang type tau-function of an arbitrary solution to 
the bigraded Toda hierarchy of $(M,1)$-type is a Dubrovin--Zhang type tau-function for the constrained KP hierarchy,
which generalizes the result in~\cite{CDZ04, FY22} for 
the Toda lattice hierarchy and the NLS hierarchy corresponding to the $M=1$ case.
\end{abstract}

%\setcounter{tocdepth}{1}
%\tableofcontents

\section{Introduction}\label{sec1}
An efficient method to the computation of logarithmic derivatives of tau-functions for integrable systems, called the {\it matrix-resolvent method}, was introduced and developed in \cite{BDY16, BDY21, CY22, D20, DVY22, DY17, DY20, DYZ21,  FY22, Y20}.
In this paper, we extend this method to the constrained KP hierarchy~\cite{C92} and the bigraded Toda hierarchy of $(M,1)$-type~\cite{C06}. 
On this basis we establish the relations between tau-functions for these two integrable hierarchies, generalizing the results in~\cite{CDZ04, FY22}.  

Let $M\ge1$ be an integer. The {\it constrained KP hierarchy}
\begin{align}\label{laxeq-cheng}
	\frac{\partial L_{\ckp}}{\partial t_{k}^{a}}=\epsilon^{-1}\Bigl[\Bigl(L_{\ckp}^{\frac{a}{M}+k}\Bigr)_{+},L_{\ckp}\Bigr],\quad k\ge 0,\, a=1,\dots,M,
\end{align}
which was introduced by Y.~Cheng~\cite{C92} (cf.~\cite{CL91, D95, KSS91, KS91}) as a reduction of the KP hierarchy (about 
the KP hierarchy see~\cite{D03,KP70}), 
is an integrable hierarchy of evolutionary partial differential equations for $(M+1)$ unknown functions, 
where 
\begin{align}\label{scl-ckp}
	L_{\ckp}:=(\epsilon\partial_{X})^{M}+v_{M-2}(\epsilon\partial_{X})^{M-2}+v_{M-3}(\epsilon\partial_{X})^{M-3}+\cdots+v_{1}\epsilon\partial_{X}+v_{0}+q(\epsilon\partial_{X})^{-1}r
\end{align}
is the Lax operator. 
Here, we recall that for a pseudo-differential operator~\cite{D03} $P=\sum_{k\in\mathbb{Z}}P_{k}\partial_{X}^{k}$, the positive part $P_{+}$ is defined as $\sum_{k\ge 0}P_{k}\partial_{X}^{k}$.  
It is easy to see that $\p/\p t^1_0=\p/\p X$, 
so below we identify $t_{0}^{1}$ with~$X$.
Several interesting properties, such as bihamiltonian structures and bilinear identities
for the constrained KP hierarchy 
have been investigated~\cite{CSZ93, CZ94, OS93}.  

Let us write the Lax operator~$L_{\ckp}$ in the matrix form
\begin{align}\label{m-lax-ckp}
	\mathcal{L}_{\ckp}(\lambda)=\epsilon\partial_{X}+\Lambda^{\ckp}(\lambda)+V_{\ckp},
\end{align}
where
\begin{align*}
&\Lambda^{\ckp}(\lambda)
=\begin{pmatrix}
		0        &   -1   &   0       &   \cdots   &  0     & 0       \\
		\vdots&\ddots    &   \ddots       &   \ddots   &  \ddots     & \vdots       \\
		\vdots&  \ddots     &   \ddots       &   \ddots   &  \ddots     & \vdots       \\
		0        & 0     &      0    &   \ddots     & -1     & 0\\
		-\lambda  &   0   &    0      &   \cdots   &  0     & 0     \\
		0        &   0   &    0      &   \cdots   &  0     & 0     
	\end{pmatrix}_{(M+1)\times (M+1)}\hspace{-5mm}, \quad 
V_{\ckp}
=\begin{pmatrix}
		0        &   0          &   \cdots   &  0     & 0     &0  \\
		\vdots        &   \vdots          &   \ddots   &  \vdots     & \vdots     &\vdots  \\
		\vdots   &\vdots      &   \ddots   &\vdots  &\vdots &\vdots\\
		0        &   0         &   \cdots   &  0     & 0     & 0\\
		v_{0}   &v_{1}     &   \cdots   &v_{M-2}& 0     & -q\\
		r        &   0         &   \cdots   &  0     & 0     & 0
\end{pmatrix}_{(M+1)\times (M+1)}.
\end{align*}
We denote by
\begin{align}
\mathcal{A}_{\ckp}:=\mathbb{C}[q_{kX},r_{kX},v_{0,kX},\dots,v_{M-2,kX}|k\ge 0]
\end{align}
the polynomial ring. Here $q_{kX}=\partial_{X}^{k}(q),r_{kX}=\partial_{X}^{k}(r),v_{l,kX}=\partial_{X}^{k}(v_{l})\, (0 \le l \le M-2), k\ge 0$. 
Introduce a gradation on $\mathcal{A}_{\ckp}[\epsilon]\otimes sl_{M+1}(\mathbb{C})((\lambda^{-1}))$ by the following degree assignments
\begin{align}
	&\deg e_{i,j}=j-i,\quad \deg \lambda=M,\quad  \deg q =-1, \quad \deg r =1,\quad \deg v_{l}=0, \quad \deg \epsilon=0, \label{deg-ckp-2}
\end{align}
where $l=0,\dots, M-2,i,j=1,\dots, M+1$, and $e_{i,j}$ denotes the $(M+1)\times (M+1)$ matrix with the $(i,j)$ entry $1$ and other entries~$0$.
The gradation defined here actually has an algebraic meaning (see Remark \ref{degree-rmk}).
Denote 
\begin{align}
\Lambda_{a}^{\ckp}(\lambda)=(-\Lambda^{\ckp}(\lambda))^{a}-\frac{M}{M+1}\lambda I\delta_{a,M},\quad 1\le a\le  M,
\end{align}
where $I$ is the $(M+1)\times (M+1)$ identity matrix. 

%We will show that
\begin{lemma}\label{lemmarnls}
There exists a unique system of solutions 
	$$(R_{1}^{\ckp}(\lambda),\dots, R_{M}^{\ckp}(\lambda))\in 
(\mathcal{A}_{\ckp}[\epsilon]\otimes sl_{M+1}(\mathbb{C})((\lambda^{-1})))^{M},$$
having the extended degrees $\deg^{e} R_{a}^{\ckp}(\lambda)=a$ (see \eqref{extend-deg-1}--\eqref{extend-deg-2}),
 to the following equations
\begin{align}
		&		[\mathcal{L}_{\ckp}(\lambda),R_{a}^{\ckp}(\lambda)]=0,\quad a=1,\dots,M,\label{reso-def-eq}\\
		&		R_{a}^{\ckp}(\lambda)=\Lambda_{a}^{\ckp}(\lambda)+{\rm lower\,\,order\,\, terms\,\, with\,\, respect\,\, to \,\, deg},\quad a=1,\dots,M,\label{ckpresolventcheng}\\
		&       \tr (R_{a}^{\ckp}(\lambda) R_{b}^{\ckp}(\lambda))=\delta_{a+b,M}(a+b)\lambda+\delta_{a+b,2M}\frac{M}{M+1}\lambda^2, \quad a,b=1,\dots,M.   \label{carconcheng}
\end{align}
\end{lemma}
The proof is in Section~\ref{sec2}. We call this unique system of solutions $R_{a}^{\ckp}(\lambda)$, $a=1,\dots,M$, the {\it basic matrix resolvents
 of $\mathcal{L}_{\ckp}(\lambda)$}.
Write 
\begin{align}
&R_{a}^{\ckp}(\lambda)=\sum_{i=0}^{M+1}\sum_{j=0}^{M+1}r_{a;i,j}^{\ckp}(\lambda)e_{i,j},\quad  r_{a;i,j}^{\ckp}(\lambda)=\sum_{k\ge -1}\frac{r_{a,k;i,j}^{\ckp}}{\lambda^{k}}.
\end{align}
For $a=1,\dots, M$, $k\ge 0$, we define the matrix-valued function
\begin{align}\label{def-ckp-V-0612}
	V_{a,k}^{\ckp}(\lambda):=\bigl(\lambda^{k}R_{a}^{\ckp}(\lambda)\bigr)_{+}-\widehat{V}_{a,k}^{\ckp},
\end{align}
where $``+"$ means taking the polynomial part in $\lambda$, and 
\begin{align}\label{core-ckp-v-0612}
	\widehat{V}_{a,k}^{\ckp}:=\sum_{i=2}^{M}\sum_{j=1}^{i-1}\sum_{l=0}^{i-j-1}\binom{l+j-1}{j-1}(\epsilon\partial_{X})^{l}(r_{a,k+1;i-j-l,M}^{\ckp})e_{i,j}.
\end{align}
We will show in Section~\ref{sec2} that the hierarchy of PDEs defined by 
\begin{align}\label{def-ckpflow}
\frac{\partial \mathcal{L}_{\ckp}(\lambda)}{\partial t_{k}^{a}}:=\epsilon^{-1}\left[V_{a,k}^{\ckp}(\lambda),\mathcal{L}_{\ckp}(\lambda)\right],\quad k\ge 0, \,\, a=1,\dots,M
\end{align}
coincides with the constrained KP hierarchy~\eqref{laxeq-cheng}. 

We introduce a sequence of elements $\Omega_{a,i;b,j}^{\ckp}\in{\mathcal A}_{\ckp}[\epsilon],\, a,b=1,\dots,M,\, i,j \geq 0$, by
\begin{align}\label{def-ckp-tau-str}
		\sum_{i,j\ge 0}\frac{\Omega_{a,i;b,j}^{\ckp}}{\lambda^{i+1}\mu^{j+1}}=\frac{\tr\left(R_{a}^{\ckp}(\lambda)R_{b}^{\ckp}(\mu)\right)}{(\lambda-\mu)^{2}}-\frac{a\lambda+b\mu}{(\lambda-\mu)^{2}}\delta_{a+b,M}-\frac{M}{M+1}\frac{\lambda \mu}{(\lambda-\mu)^{2}}\delta_{a+b,2M}.
\end{align}
To see $\Omega_{a,i;b,j}^{\ckp}$ are well defined see Section~\ref{sec2}. We also prove in Section~\ref{sec2} the following lemma. 
\begin{lemma}\label{tau-str-ckp}
The elements $\Omega_{a,k;b,\ell}^{\ckp}, 1\le a,b\le M,k,\ell\ge 0$, have the following properties:
\begin{align}
		&  \Omega_{a,k;b,\ell}^{\ckp}  =\Omega_{b,\ell;a,k}^{\ckp},\quad \forall \, a,b=1,\dots,M,\,k,\ell\geq 0, \\
		& \p_{t_{i}^{c}} \Omega_{a,k;b,\ell}^{\ckp}   = \p_{t_{k}^{a}} \Omega_{b,\ell;c,i}^{\ckp}= 
		\p_{t^{b}_{\ell}} \Omega_{c,i;a,k}^{\ckp}, \quad \forall \, a,b,c=1,\dots,M,\,k,\ell,i\geq 0.
\end{align}
\end{lemma}
We call $\Omega_{a,i;b,j}^{\ckp},1\le a,b\le M,i,j\ge 0$ the {\it two-point correlation functions} ({\it aka} the {\it tau-structure}~\cite{CDZ04, DZ, DZ04}) for 
the constrained KP hierarchy. 
For $k\ge 3$, define the $k$-point correlation functions by
\begin{align}
	\Omega_{a_{1},i_{1};\dots;a_{k},i_{k}}^{\ckp}:=\epsilon^{k-2}\frac{\partial^{k-2}\Omega_{a_{k-1},i_{k-1};a_{k},i_{k}}^{\ckp}}{\partial t^{a_{1}}_{i_{1}}\cdots \partial t^{a_{k-2}}_{i_{k-2}}},\quad 1\le a_{1},\dots,a_{k}\le M,i_{1},\dots,i_{k}\ge 0.
\end{align}

\begin{prop}\label{main-thm}
For any integer $k\ge 2$, and any fixed $a_1,\dots,a_k\in \{1,\dots,M\}$, we have
\begin{align}
		\sum_{i_{1},\dots,i_{k}\ge 0}\frac{\Omega_{a_{1},i_{1};\dots;a_{k},i_{k}}^{\ckp}}{\prod_{j=1}^{k}\lambda_{j}^{i_{k}+1}}
		=&-\sum_{\sigma\in S_{k}/C_{k}}\frac{\tr \bigl(R_{a_{\sigma(1)}}^{\ckp}(\lambda_{\sigma(1)}) \cdots R_{a_{\sigma(k)}}^{\ckp}(\lambda_{\sigma(k)})\bigr)}{\prod_{i=1}^{k} (\lambda_{\sigma(i)}-\lambda_{\sigma(i+1)})} \nonumber\\
		&-\left(\frac{a_{1}\lambda_{1}+a_{2}\lambda_{2}}{(\lambda_{1}-\lambda_{2})^{2}}\delta_{a_{1}+a_{2},M}+\frac{M}{M+1}\frac{\lambda_{1} \lambda_{2}}{(\lambda_{1}-\lambda_{2})^{2}}\delta_{a_{1}+a_{2},2M}\right)\delta_{k,2},
\end{align}
where $S_{k}$ denotes the symmetry group, $C_{k}$ denotes the cyclic group, and it is understood that $\sigma(k+1)=\sigma(1)$.
\end{prop}
Let 
\begin{align}\label{solution-ckp-cheng-sec1}
	\left(q({\bf t};\epsilon),r({\bf t};\epsilon),v_{0}({\bf t};\epsilon),\dots,v_{M-2}({\bf t};\epsilon)\right),\quad {\bf t}=(t^{a}_{k})_{k\ge 0}^{a=1,\dots,M},
\end{align}	
be a solution to the constrained KP hierarchy~\eqref{def-ckpflow}. 
For $k\ge 2$, write $\Omega_{a_{1},i_{1};\dots;a_{k},i_{k}}^{\ckp}({\bf t};\epsilon)$ as the $k$-point correlation functions 
evaluated at the solution~\eqref{solution-ckp-cheng-sec1}.
It follows from Lemma~\ref{tau-str-ckp} that there exists a function $\tau_{\ckp}({\bf t};\epsilon)$, such that
\begin{align}\label{def-tau-func-ckp}
	\epsilon^{2}\frac{\partial^{2}\log \tau_{\ckp}({\bf t};\epsilon)}{\partial {t_{i}^{a}}\partial {t_{j}^{b}}}=\Omega_{a,i;b,j}^{\ckp}({\bf t};\epsilon),\quad i,j\ge 0,1\le a,b\le M.
\end{align}
We call $\tau_{\ckp}({\bf t};\epsilon)$ the {\it Dubrovin--Zhang type tau-function} of the solution~\eqref{solution-ckp-cheng-sec1} 
to the constrained KP hierarchy~\eqref{def-ckpflow}. 
The function $\tau_{\ckp}({\bf t};\epsilon)$ is determined uniquely by~\eqref{def-tau-func-ckp}
up to multiplying by the exponential of a linear function
\begin{align}
	\tau_{\ckp}({\bf t};\epsilon)\mapsto e^{a_{0}+\sum_{i=1}^{M}\sum_{k=0}^{\infty}a_{i,k}t_{k}^{i}}\tau_{\ckp}({\bf t};\epsilon),\quad a_{0},a_{i,k}\in \mathbb{C}((\epsilon)).
\end{align}

In~\cite{LZZ}, S.-Q. Liu, Y.~Zhang and X.~Zhou studied the constrained KP hierarchy in suitable variables from the 
viewpoints of bihamiltonian systems and semisimple Frobenius manifolds. In particular, the dispersionless limit of the 
constrained KP hierarchy and the corresponding Frobenius manifold were given in~\cite{Du96, LZZ}, and the central invariants for the bihamiltonian structure 
of the constrained KP hierarchy were shown to be all equal to 1/24. Based on these results, Liu, Zhang and Zhou~\cite{LZZ} conjectured that 
the constrained KP hierarchy is topological, namely, it is equivalent to the Dubrovin--Zhang hierarchy 
of the Frobenius manifold. In~\cite{LWZ21}, S.-Q. Liu, Z.~Wang and Y.~Zhang prove that the Dubrovin--Zhang hierarchy of 
a semisimple Frobenius manifold has a polynomial bihamiltonian structure (the central invariants for the bihamiltonian structure of 
the Dubrovin--Zhang hierarchy are known to be all equal to 1/24~\cite{DLZ08}). This result together with the results in~\cite{LZZ} 
confirms the Liu--Zhang--Zhou conjecture; a more direct proof of the conjecture of Liu--Zhang--Zhou is given in~\cite{CLPS22}. 

When $M=1$, the constrained KP hierarchy becomes the celebrated nonlinear Schr\"odinger
(NLS) hierarchy ({\it aka} the {\it AKNS hierarchy}) (cf.~\cite{AC91, AS, C92, CL91, D81-1, D81-2, D95, D20, D03, FY22, KSS91, KS91, Taka, ZS72}). 
This hierarchy is connected to the Toda lattice hierarchy~\cite{CDZ04, F74, svM74, UT} by exchanging the role of the space and time variables~\cite{CDZ04} 
(see also~\cite{FY22} for more details). It is natural to consider the generalization of this result to the case when $M\ge2$. 
To this end, let us consider the bigraded Toda hierarchy~\cite{C06}. 
Let $L$ denote the following difference Lax operator
\begin{align}\label{scl-bt}
	L=\mathcal{T}^{M}+u_{M-1}\mathcal{T}^{M-1}+u_{M-2}\mathcal{T}^{M-2}+\cdots+u_{0}+u_{-1}\mathcal{T}^{-1},
\end{align}
where $\mathcal{T}=e^{\epsilon\partial_{x}}$ is the shift operator. 
The {\it bigraded Toda hierarchy of $(M,1)$-type} are defined~\cite{C06} as the following Lax equations
\begin{align}\label{laxeq-bt}
	\frac{\partial L}{\partial t_{k}^{a}}=\epsilon^{-1}\Bigl[\bigl(L^{\frac{a}{M}+k}\bigr)_{+},L\Bigr],\quad k\ge 0, \, a=1,\dots,M.
\end{align}
Here, for any difference operator $P$ of the form $P=\sum_{k\in \mathbb{Z}}P_{k}\mathcal{T}^{k}$, its positive part $P_{+}$ is defined by $P=\sum_{k\ge 0}P_{k}\mathcal{T}^{k}$.
The matrix form of the Lax operator $L$ is
\begin{align}
	\mathcal{L}(\lambda)=\mathcal{T}+\Lambda(\lambda)+V,
\end{align}
where 
\begin{align}\label{def-lam-V}
\Lambda(\lambda)
=\begin{pmatrix}
		0           &   \cdots       &   0  &  -\lambda     & 0     \\
		-1    &    \ddots     &  \ddots    &  0     & 0      \\
		0   &   \ddots       &  \ddots    &  \vdots     & \vdots      \\
		\vdots            & \ddots         &   \ddots     & 0     & 0\\
		0  &    \cdots    &            0     &   -1      &0
\end{pmatrix}_{(M+1)\times (M+1)} \hspace{-5mm}, \quad 
 V
 =\begin{pmatrix}
		u_{M-1}  &u_{M-2} &\cdots &u_{0} &u_{-1}\\
		0                      &0&\cdots  & 0      &  0   \\
		\vdots        &\vdots&\ddots    & \vdots         & \vdots\\
		\vdots            &     \vdots          &    \ddots & \vdots           &\vdots  \\
		0             & 0 &    \cdots & 0           &0 
	\end{pmatrix}_{(M+1)\times (M+1)}.
\end{align}
We denote by
\begin{align}
		\mathcal{A}:=\mathbb{C}[u_{l,kx}|-1\le l\le M-1,k\ge 0]
\end{align}
the polynomial ring. Here $u_{l,kx}=\partial_{x}^{k}(u_{l})\, (-1\le l\le M-1), k\ge 0$.
Introduce a gradation on $\mat(M+1,\mathbb{C}((\lambda^{-1})))$ by the following degree assignments
\begin{align}\label{bt-degree}
	\overline{\deg}\, \lambda=M,\quad \overline{\deg}\, e_{i,j}=i-j,\quad i,j=1,\dots,M+1,
\end{align}
and denote $\Lambda_{a}(\lambda):=(-\Lambda(\lambda))^{a}$. We will show that

\begin{lemma}\label{bt-MR-def}
There exists a unique system of solutions 
	$$(R_{1}(\lambda),\dots, R_{M}(\lambda))\in 
	(\mathcal{A}[[\epsilon]]\otimes \mat(M+1,\mathbb{C}((\lambda^{-1}))))^{M},$$ 
	having the extended degrees $\overline{\deg}^{e} R_{a}(\lambda)=a$ (see \eqref{bt-extend-deg-1}--\eqref{bt-extend-deg-2}),
	to the following equations
\begin{align}
		&		\mathcal{T}(R_{a}(\lambda) )(\Lambda(\lambda)+V)-(\Lambda(\lambda)+V)R_{a}(\lambda) =0, \quad a=1,\dots,M,\label{defbtodreso}\\
		&		R_{a}(\lambda)=\Lambda_{a}(\lambda)+ {\rm lower\,\,order\,\, terms\,\, with\,\, respect\,\, to \,\, \overline{deg}},\quad a=1,\dots,M,\label{reso}\\		
		&      \tr(R_{a}(\lambda)R_{b}(\lambda))=\delta_{a+b,M}(a+b)\lambda+M\lambda^{2}\delta_{a+b,2M},\quad 
		\tr R_{a}(\lambda)=M\delta_{a,M}\lambda,\quad  a,b=1,\dots, M.\label{carconchengbt}
\end{align}
\end{lemma}
We call this unique system of solutions $R_{a}(\lambda)$, $a=1,\dots,M$, the 
{\it basic matrix resolvents of $\mathcal{L}(\lambda)$}.
Write 
\begin{align}
&R_{a}(\lambda)=\sum_{i=0}^{M+1}\sum_{j=0}^{M+1}r_{a;i,j}(\lambda)e_{i,j},\quad r_{a;i,j}(\lambda)=\sum_{k\ge -1}\frac{r_{a,k;i,j}}{\lambda^{k}}.
\end{align}
For any $a=1,\dots, M, k\ge 0$, we define the matrix-valued function
\begin{align}\label{def-ckp-V}
	V_{a,k}(\lambda):=\left(\lambda^{k}R_{a}(\lambda)\right)_{+}-\widehat{V}_{a,k},
\end{align}
where 
\begin{align}\label{m-bt}
	\widehat{V}_{a,k}:=-r_{a,k+1;M+1,1}e_{M+1,M+1}+\sum_{i=1}^{M-1}\sum_{j=i+1}^{M}\mathcal{T}^{M+1-j}(r_{a,k+1;M+1-j+i,1})e_{i,j}.
\end{align}
We will show in Section~\ref{sec3} that  the hierarchy of PDEs defined by 
\begin{align}\label{def-btflow}
	\epsilon\frac{\partial \mathcal{L}(\lambda)}{\partial t_{k}^{a}}:=\mathcal{T}(V_{a,k}(\lambda))(\Lambda(\lambda)+V)-(\Lambda(\lambda)+V)V_{a,k}(\lambda),\quad k\ge 0, 1\le a\le M,
\end{align}
coincides with the bigraded Toda hierarchy of $(M,1)$-type~\eqref{laxeq-bt}. 

Like above, let us  introduce a sequence of elements $\Omega_{a,i;b,j}\in{\mathcal A}[[\epsilon]],\, a,b=1,\dots,M,\, i,j \geq 0$, by
\begin{align}\label{def-bt-tau-str}
		\sum_{i,j\ge 0}\frac{\Omega_{a,i;b,j}}{\lambda^{i+1}\mu^{j+1}}=\frac{\tr(R_{a}(\lambda)R_{b}(\mu))}{(\lambda-\mu)^{2}}-\frac{a\lambda+b\mu}{(\lambda-\mu)^{2}}\delta_{a+b,M}-M\frac{\lambda \mu}{(\lambda-\mu)^{2}}\delta_{a+b,2M}.
\end{align}
\begin{lemma}\label{tau-str-bt}
The elements $\Omega_{a,k;b,\ell}$ have the following properties:
\begin{align}
	&\Omega_{a,k;b,\ell}  =\Omega_{b,\ell;a,k},\quad \forall \, a,b=1,\dots,M,\,k,\ell\geq 0, \\
		& \p_{t_{j}^{c}} \Omega_{a,k;b,\ell}   = \p_{t_{k}^{a}} \Omega_{b,\ell;c,j}= 
		\p_{t_{\ell}^{b}} \Omega_{c,j;a,k}, \quad \forall \, a,b,c=1,\dots,M,\,k,\ell,j\geq 0.
\end{align}
\end{lemma}
The proof is in Section~\ref{sec3}.
We call $\Omega_{a,i;b,j},1\le a,b\le M,i,j\ge 0$ the {\it two-point correlation functions} ({\it aka} the {\it tau-structure}~\cite{CDZ04, DZ, DZ04}) for the bigraded Toda hierarchy of $(M,1)$-type. 
For $k\ge 3$, define the $k$-point correlation functions by
\begin{align}
	\Omega_{a_{1},i_{1};\dots;a_{k},i_{k}}:=
	\epsilon^{k-2}\frac{\partial^{k-2}\Omega_{a_{k-1},i_{k-1};a_{k},i_{k}}}{\partial t^{a_{1}}_{i_{1}}\cdots \partial t^{a_{k-2}}_{i_{k-2}}}, 
	\quad 1\le a_{1},\dots,a_{k}\le M, \, i_{1},\dots,i_{k}\ge 0.
\end{align}
\begin{prop}\label{main-thm-bt}
For any $k\ge 2$, and any fixed $a_1,\dots,a_k\in \{1,\dots,M\}$, we have
\begin{align}
		\sum_{i_{1},\dots,i_{k}\ge 0}\frac{\Omega_{a_{1},i_{1};\dots;a_{k},i_{k}}}{\prod_{j=1}^{k}\lambda_{j}^{i_{k}+1}}
		=&-\sum_{\sigma\in S_{k}/C_{k}}\frac{\tr \bigl(R_{a_{\sigma(1)}}(\lambda_{\sigma(1)})\cdots R_{a_{\sigma(k)}}(\lambda_{\sigma(k)})\bigr)}{\prod_{i=1}^{k}(\lambda_{\sigma(i)}-\lambda_{\sigma(i+1)})} \nonumber\\
		&-\left(\frac{a_{1}\lambda_{1}+a_{2}\lambda_{2}}{(\lambda_{1}-\lambda_{2})^{2}}\delta_{a_{1}+a_{2},M}+M\frac{\lambda_{1} \lambda_{2}}{(\lambda_{1}-\lambda_{2})^{2}}\delta_{a_{1}+a_{2},2M}\right)\delta_{k,2}.
\end{align}
\end{prop}
Let 
\begin{align}\label{solution-bt-sec1}
	\left(u_{-1}(x,{\bf t};\epsilon),u_{0}(x,{\bf t};\epsilon),\dots,u_{M-1}(x,{\bf t};\epsilon)\right)
\end{align}	
be a solution to the bigraded Toda hierarchy~\eqref{def-btflow}. Let $R_{a}({\lambda;x,{\bf t}};\epsilon)$ denote the $R_{a}(\lambda)$ evaluated at the solution~\eqref{solution-bt-sec1}.  We will show in Section~\ref{sec3} the following lemma.
\begin{lemma}\label{defbtodataufun}
There exists a function $\tau(x,{\bf t};\epsilon)$, such that
\begin{align}
		&	\sum_{i,j\ge 0}\frac{\epsilon^{2}\frac{\partial^{2} \log \tau(x,{\bf t};\epsilon)}{\partial t_{i}^{a} \partial t_{j}^{b}}}{\lambda^{i+1}\mu^{j+1}}=\frac{\tr(R_{a}(\lambda;x,{\bf t};\epsilon)R_{b}(\mu;x,{\bf t};\epsilon))}{(\lambda-\mu)^{2}}-\frac{a\lambda+b\mu}{(\lambda-\mu)^{2}}\delta_{a+b,M}-M\frac{\lambda \mu}{(\lambda-\mu)^{2}}\delta_{a+b,2M},\label{def-tau-1}\\
		&  \delta_{a,M}+\sum_{i\ge 0}\frac{\epsilon}{\lambda^{i+1}}\frac{\partial}{\partial t_{i}^{a}}\log \frac{\tau(x+\epsilon,{\bf t};\epsilon)}{\tau(x,{\bf t};\epsilon)}=\bigl[R_{a}(\lambda;x+\epsilon,{\bf t};\epsilon)\bigr]_{M+1,1},\label{def-tau-2}\\
		&  \frac{\tau(x+\epsilon,{\bf t};\epsilon)\tau(x-\epsilon,{\bf t};\epsilon)}{\tau(x,{\bf t};\epsilon)^{2}}=u_{-1}(x,{\bf t};\epsilon)  \label{def-tau-3}.
\end{align}
\end{lemma}

The function $\tau(x,{\bf t};\epsilon)$ is determined uniquely by~\eqref{def-tau-1}--\eqref{def-tau-3} up to
\begin{align}
		\tau(x,{\bf t};\epsilon)\mapsto e^{a_{0}+a_{1}x+\sum_{a=1}^{M}\sum_{j=0}^{\infty}b_{j}^{a}t_{j}^{a}}\tau(x,{\bf t};\epsilon),\quad a_{0},a_{1},b_{j}^{a}\in\mathbb{C}((\lambda^{-1})).
\end{align}
We call $\tau(x,{\bf t};\epsilon)$ the {\it Dubrovin--Zhang type tau-function} of the solution~\eqref{solution-bt-sec1} 
to the bigraded Toda hierarchy of $(M,1)$-type \eqref{def-btflow}.

We note that Proposition~\ref{main-thm-bt} can be used to compute the orbifold Gromov-Witten invariants of $\mathbb{P}^{1}$-orbifolds~\cite{MT08, Rossi}.

It is known that the Frobenius manifold $M_{\rm cKP}$ corresponding to the constrained KP hierarchy 
is related by a Legendre-type transformation (cf.~\cite{Du96, LZZ})
to the Frobenius manifold $M_{\rm bToda}$ corresponding to the bigraded Toda hierarchy of $(M,1)$-type. Moreover, Bakalov and Wheeles showed~\cite{BW-16} that the bigraded Toda hierarchy 
admits the Virasoro symmetries which has the same expressions as the ones of the Frobenius manifold $M_{\rm bToda}$, which 
indicates that the bigraded Toda hierarchy is topological (see also~\cite{BW-16, Rossi} for a different argument).
For the case of $M=1$, the bigraded Toda hierarchy~\eqref{laxeq-bt} is just the Toda lattice hierarchy, 
which is equivalent to the constrained KP hierarchy~\eqref{laxeq-cheng} with $M=1$ (the NLS hierarchy),
namely, for any solution $(u_{-1}, u_{0})$ to the Toda lattice hierarchy, the functions $q,r$ defined by
\begin{align}\label{qr-relation}
	q=\frac{\mathcal{T}(\tau)}{\tau},\quad r=\frac{\mathcal{T}^{-1}(\tau)}{\tau}
\end{align}
satisfy the NLS hierarchy,   
as it was shown in~\cite{CDZ04} (see also~\cite{FY22}), 
where we recall that $\mathcal{T}=e^{\epsilon\partial_{x}}$ is the shift operator and $\tau$ is the tau-function of the solution $(u_{-1},u_{0})$ to 
the Toda lattice hierarchy (alternatively, we have $qr=u_{-1},\epsilon\frac{q_{X}}{q}=u_{0}$); for 
the more precise meaning of this equivalence see~\cite{CDZ04, FY22} and the materials below given in this paper. 
Motivated by this equivalence, by Liu--Zhang--Zhou's conjecture, by the result of 
Bakalov--Wheeles \cite{BW-16}  and by the result of Rossi \cite{Rossi}, 
we prove in this paper the following theorem generalizing this equivalence from $M=1$ to arbitrary $M\ge 1$.

\begin{theorem}\label{maintheorem}
Let $(u_{-1},\dots,u_{M-1})$ 
be an arbitrary solution to the bigraded Toda hierarchy of $(M,1)$-type~\eqref{def-btflow}, 
and $\tau$ the tau-function of this solution. Then we have

1. The vector $(q,r,v_0, \dots, v_{M-2})$ 
defined by~\eqref{keylaxrelation} 
is a solution to the constrained KP hierarchy~\eqref{def-ckpflow}; 

2. For any~$x$, the function 
 $\tau$ is the tau-function of $(q,r,v_0, \dots, v_{M-2})$ to the constrained KP hierarchy.
\end{theorem}
We note that the results similar to the statements of the above theorem were also obtained by 
Carlet, van de Leur, Posthuma and Shadrin~\cite{CLPS21,CLPS22} by a different approach; however, the map~\eqref{keylaxrelation} 
relating $(u_{-1},\dots,u_{M-1})$ and $(q,r,v_0, \dots, v_{M-2})$ is not 
given explicitly there.    

Let us present the first statement of Theorem~\ref{maintheorem} more explicitly by means of examples. 
For $M=2$, Theorem~\ref{maintheorem} says that, for any solution $(u_{-1},u_{0},u_{1})$ to the bigraded Toda hierarchy of $(2,1)$-type, the functions $q,r,v_{0}$, defined by
\begin{align}
&q =\frac{\mathcal{T}(\tau)}{\tau},\quad r=\frac{\mathcal{T}^{-1}(\tau)}{\tau},\label{qr-def-603}\\
&v_{0}  =-\bigl((1+\mathcal{T})^{-1}(u_{1})\bigr)^{2}-\epsilon (1+\mathcal{T})^{-1}(u_{1,X})+u_{0}\label{v0-def-603}
\end{align}
satisfy the constrained KP hierarchy.  

For $M=3$, Theorem~\ref{maintheorem} says that, 
for any solution $(u_{-1},u_{0},u_{1},u_{2})$ to the bigraded Toda hierarchy of $(3,1)$-type, the functions $q,r,v_{0},v_{1}$ defined by
\begin{align}
&q=\frac{\mathcal{T}(\tau)}{\tau},\quad r=\frac{\mathcal{T}^{-1}(\tau)}{\tau},\label{qr-def-1}\\
&v_{0}=u_{0}+\alpha(\epsilon(\mathcal{T}-1)(\alpha_{X})-u_{1}+\mathcal{T}(\alpha)(1+\mathcal{T})(\alpha))-\epsilon^{2}\alpha_{XX} \label{v0-def-1}\\
&v_{1}=u_{1}-2 \epsilon \alpha_{X}-\epsilon \mathcal{T}(\alpha_{X})-\mathcal{T}(\alpha)(1+\mathcal{T})(\alpha)-\alpha^{2} 
\end{align}
satisfy the constrained KP hierarchy, 
where 
\begin{align}
\alpha:=(1+\mathcal{T}+\mathcal{T}^{2})^{-1}(u_{2}).
\end{align}

The converse statement of Theorem~\ref{maintheorem} is essentially also true 
(see~\eqref{xderiv} and the materials after of Section~\ref{sec4}). 

We hope that the method used in this paper can be helpful for the study 
of the topological deformations of the principal hierarchies  
of the Frobenius manifolds associated to the extended affine Weyl groups of $BCD$ type constructed in~\cite{DSZZ19, DZ98, Z20}.

%%%%%%%%%%%%%%%%%%%%%%%%%%%%%%%%%%%%%%%%%%

The paper is organized as follows. In Sections~\ref{sec2}, \ref{sec3}, we study tau-functions for the constrained KP hierarchy 
 and the bigraded Toda hierarchy of $(M,1)$-type. In Section~\ref{sec4}, we prove Theorem~\ref{maintheorem}. 

\medskip

\noindent {\bf Acknowledgements} We would like to thank Prof. Yi Cheng and Prof. Youjin Zhang for their advice and helpful discussions. The work is partially supported by NSFC (No.12071451, No.12061131014).

\section{The matrix-resolvent method to tau-functions for the  constrained KP hierarchy}\label{sec2}	

\subsection{Matrix resolvents for the constrained KP hierarchy}
Let ${\rm ad}:sl_{M+1}(\mathbb{C})\to gl(sl_{M+1}(\mathbb{C}))$ be the adjoint representation of $sl_{M+1}(\mathbb{C})$. Then we have
\begin{lemma}\label{ker-im-ckp}
The spaces $\im\ad_{\Lambda^{\ckp}(\lambda)},\ke\ad_{\Lambda^{\ckp}(\lambda)}\subset sl_{M+1}(\mathbb{C})((\lambda^{-1}))$ have the following form
\begin{align}\label{ckp-im-ker}
&\im\ad_{\Lambda^{\ckp}(\lambda)}=\left\{\sum_{l=1}^{M^{2}+M}\alpha_{l}(\lambda)e_{l}^{\ckp}(\lambda)\bigg|\alpha_{l}(\lambda)\in \mathbb{C}((\lambda^{-1}))\right\},\\
&\ke\ad_{\Lambda^{\ckp}(\lambda)}=\left\{\sum_{a=1}^{M}\beta_{a}(\lambda)\Lambda_{a}^{\ckp}(\lambda)\bigg|\beta_{a}(\lambda)\in\mathbb{C}((\lambda^{-1}))\right\}=:\mathfrak{h}_{\ckp}.
\end{align}
Here,
\begin{align*}
		&e_{k(M-1)+i}^{\ckp}(\lambda)=\left\{\begin{aligned}
			&	 e_{1,k+1}- e_{i+1,i+k+1}\quad  & i=1,\dots ,M-k-1\\
			&	 e_{1,k+1}-\lambda e_{i+1,i+k+1-M}\quad       & i=M-k,\dots, M-1
		\end{aligned}  
		\right. ,\quad k=0,\dots,M-1,\\
		&e_{M(M-1)+a}^{\ckp}(\lambda)= e_{M+1,a},\quad e_{(M+1)(M-1)+a+1}^{\ckp}(\lambda)=e_{a,M+1},\quad a=1,\dots, M.
\end{align*}
\end{lemma}
\noindent  The proof is given in Appendix~\ref{appen-A}.  

Note that
$e_{l}^{\ckp}(\lambda),\Lambda_{a}^{\ckp}(\lambda),1\le l\le M^{2}+M, 1\le a\le M$, are linearly independent. We have 
\begin{align}
		\dim_{\mathbb{C}((\lambda^{-1}))} \ke \ad_{\Lambda^{\ckp}(\lambda)}+\dim_{\mathbb{C}((\lambda^{-1}))} \im \ad_{\Lambda^{\ckp}(\lambda)}=(M+1)^{2}-1=\dim_{\mathbb{C}((\lambda^{-1}))} sl_{M+1}(\mathbb{C})((\lambda^{-1})).
\end{align}
Combining with
$\ke \ad_{\Lambda^{\ckp}(\lambda)} \cap \im \ad_{\Lambda^{\ckp}(\lambda)}=0$,
we obtain the following lemma.
\begin{lemma}\label{decomuni}
The linear space	$sl_{M+1}(\mathbb{C})((\lambda^{-1}))$ admits the decomposition
\begin{align}\label{deco-ckp}
		sl_{M+1}(\mathbb{C})((\lambda^{-1}))=\ke \ad_{\Lambda^{\ckp}(\lambda)} \oplus \im \ad_{\Lambda^{\ckp}(\lambda)}.
\end{align}
\end{lemma}
The following lemma generalizes the result of Drinfeld and Sokolov~\cite{DS85} (cf.~\cite{BDY21, DYZ21}).
\begin{lemma}\label{funlemma}
Let $	\mathcal{L}_{\ckp}(\lambda)=\epsilon\partial_{X}+\Lambda^{\ckp}(\lambda)+V_{\ckp} $. 
Then there exists a unique pair $(U_{\ckp} ,H_{\ckp} )$ of the form
\begin{align}
  &U_{\ckp} =\sum_{k\ge 1}U_{\ckp}^{[-k]}\in \mathcal{A}_{\ckp} [\epsilon]\otimes\im \ad_{\Lambda^{\ckp}(\lambda)},\label{U-form-ckp}\\
  &H_{\ckp} =\sum_{k\ge 0}H_{\ckp}^{[-k]}\in \mathcal{A}_{\ckp} [\epsilon]\otimes\ke \ad_{\Lambda^{\ckp}(\lambda)},\label{H-form-ckp}
\end{align}
where $\im, \ke $ are taken in $sl_{M+1}(\mathbb{C})((\lambda^{-1}))$, such that 
\begin{align}\label{flemma}
	e^{-\ad_{U_{\ckp} } }(\mathcal{L}_{\ckp}(\lambda))=\epsilon\partial_{X}+\Lambda^{\ckp}(\lambda)+H_{\ckp}.
\end{align}
Here $U_{\ckp}^{[-k]},H_{\ckp}^{[-k]}$ denote the elements in $\mathcal{A}_{\ckp} [\epsilon]\otimes sl_{M+1}(\mathbb{C})((\lambda^{-1}))$ with degree $-k$.
\end{lemma}
\begin{proof} 
Note that equation \eqref{flemma} is equivalent to
\begin{align}\label{recurrUH}
	\sum_{k=1}^{\infty}\frac{(-1)^{k-1}}{k!}\ad_{U_{\ckp}}^{k-1}(\epsilon\partial_{X}( U_{\ckp}))+V_{\ckp}+\sum_{k=1}^{\infty}\frac{(-1)^{k}}{k!}\ad_{U_{\ckp}}^{k}(\Lambda^{\ckp}(\lambda)+V_{\ckp})=H_{\ckp}.
\end{align}
Substituting~\eqref{U-form-ckp} and~\eqref{H-form-ckp} into~\eqref{recurrUH} and 
comparing the components of both sides of~\eqref{recurrUH}  with degree $-k$, we find that \eqref{recurrUH} is equivalent to
\begin{align} \label{recur}
	H_{\ckp}^{[-k]} +\left[U_{\ckp}^{[-k-1]},\Lambda^{\ckp}(\lambda)\right]=G_{k}^{\ckp}, \quad k\ge 0,
\end{align}
where $G_{k}^{\ckp}\in \mathcal{A}_{\ckp}[\epsilon]\otimes sl_{M+1}(\mathbb{C})((\lambda^{-1}))$. Here $G_{k}^{\ckp}$ are determined by $V_{\ckp},U_{\ckp}^{[-1]},\dots,U_{\ckp}^{[-k]}$.
We will prove the lemma by induction on the degrees.
 For $k=0$ equation \eqref{recur} reads
\begin{align}
		H_{\ckp}^{[0]}+\left[U_{\ckp}^{[-1]},\Lambda^{\ckp}(\lambda)\right]=V_{\ckp}^{[0]}.\label{recur0}
\end{align}
Decompose
\begin{align}
V_{\ckp}^{[0]}=V_{\ckp}^{[0],\im}+V_{\ckp}^{[0],\ke},
\end{align}
where $V_{\ckp}^{[0],\im}\in \mathcal{A}_{\ckp}[\epsilon]\otimes \im \ad_{\Lambda^{\ckp}(\lambda)}, V_{\ckp}^{[0],\ke}\in \mathcal{A}_{\ckp}[\epsilon]\otimes \ke \ad_{\Lambda^{\ckp}(\lambda)}$.
Then by using Lemma~\ref{decomuni}, 
equation~\eqref{recur0} is equivalent to
\begin{align}
		H_{\ckp}^{[0]}=V_{\ckp}^{[0],\ke},\quad \left[U_{\ckp}^{[-1]},\Lambda^{\ckp}(\lambda)\right]=V_{\ckp}^{[0],\im}.
\end{align}
It is easy to see that the map $\ad_{\Lambda^{\ckp}(\lambda)}: \im \ad_{\Lambda^{\ckp}(\lambda)}\to \im \ad_{\Lambda^{\ckp}(\lambda)}$ is invertible, 
then we have 
\begin{align}
		U_{\ckp}^{[-1]}=-\ad_{\Lambda^{\ckp}(\lambda)}^{-1}(V_{\ckp}^{[0],\im}).
\end{align}
The second step of the induction follows from equation \eqref{recur} and Lemma \ref{decomuni}. 
The lemma is proved.
\end{proof}

\begin{remark}\label{degree-rmk}
Following~\cite{ LZZ} (cf.~\cite{B-65,CDZ04, DR1906, H-72, OS93}), define 
\begin{align}\label{v-1m-1}
v_{-1}:=qr, \quad v_{M-1}:=\epsilon\frac{q_{X}}{q}.
\end{align}
Let $\widetilde{G}= {\rm diag} (1,\dots,1,q)$. 
%\begin{pmatrix}
%	1 &               &    & \\
%	&\ddots   &    & \\
%	&                &1 &  \\
%	&                &   &q
%\end{pmatrix}$. 
Consider the following {\it gauge transformation} $V\mapsto \widetilde{V}$ given by
\begin{align}\label{gauge}
	\widetilde{G}\circ \mathcal{L}_{\ckp} \circ \widetilde{G}^{-1}=\epsilon\partial_{X}+\widetilde{\Lambda}(\lambda)+\widetilde{V}=:\widetilde{\mathcal{L}},
\end{align}
where
\begin{align}
	\widetilde{\Lambda}(\lambda)=\begin{pmatrix}
		0        &   -1   &   0       &   \cdots   &  0     & 0       \\
		\vdots&\ddots    &   \ddots       &   \ddots   &  \ddots     & \vdots       \\
		\vdots&  \ddots     &   \ddots       &   \ddots   &  \ddots     & \vdots       \\
		0        & 0     &      0    &   \ddots     & -1     & 0\\
		-\lambda  &   0   &    0      &   \cdots   &  0     & -1     \\
		0        &   0   &    0      &   \cdots   &  0     & 0     
	\end{pmatrix},\quad \widetilde{V}=\begin{pmatrix}
	0        &   0          &   \cdots   &  0     & 0     &0  \\
	\vdots        &   \vdots          &   \ddots   &  \vdots     & \vdots     &\vdots  \\
	\vdots   &\vdots      &   \ddots   &\vdots  &\vdots &\vdots\\
	0        &   0         &   \cdots   &  0     & 0     & 0\\
	v_{0}   &v_{1}     &   \cdots   &v_{M-2}& 0     & 0\\
	v_{-1}        &   0         &   \cdots   &  0     & 0     & -v_{M-1}
\end{pmatrix}.
\end{align}
The matrix Lax operator $\widetilde{\mathcal{L}}$ corresponds to the following scalar Lax operator
\begin{align}
	\widetilde{L}=(\epsilon\partial_{X})^{M}+v_{M-2}(\epsilon\partial_{X})^{M-2}+v_{M-3}(\epsilon\partial_{X})^{M-3}+\cdots+v_{1}\epsilon\partial_{X}+v_{0}+(\epsilon\partial_{X}-v_{M-1})^{-1}v_{-1},
\end{align}	
which is the Lax operator used in~\cite{LZZ, OS93}. Obviously, under the degree assignments~\eqref{deg-ckp-2}, 
 $\deg v_{l}=0$, $l=-1,\dots, M-1$. This explains that the gradation given by $\deg$ indeed has an algebraic meaning.
From this viewpoint, the above proof is essentially the same as that in~\cite{BDY21,DS85, DVY22}.
\end{remark}

\begin{example}
Comparing the degree $-1$ parts of both sides of equation~\eqref{recurrUH},
	we obtain
	\begin{align}\label{exam-bt-2}
		H_{\ckp}^{[-1]} +\left[U_{\ckp}^{[-2]},\Lambda^{\ckp}(\lambda)\right]=&\epsilon \partial_{X}(U_{\ckp}^{[-1]})+V_{\ckp}^{[-1]}-\left[U_{\ckp}^{[-1]},V_{\ckp}^{[0]}\right] +\frac{1}{2}\left[U_{\ckp}^{[-1]},\left[U_{\ckp}^{[-1]},\Lambda^{\ckp}(\lambda)\right]\right].
	\end{align}
	Since $U_{\ckp}^{[-2]}\in \mathcal{A}_{\ckp}[\epsilon]\otimes \im \ad_{\Lambda^{\ckp}(\lambda)},H_{\ckp}^{[-1]}\in \mathcal{A}_{\ckp}[\epsilon]\otimes \ke \ad_{\Lambda^{\ckp}(\lambda)}$ and the right-hand side of~\eqref{exam-bt-2} belongs to $\mathcal{A}_{\ckp}[\epsilon]\otimes sl_{M+1}(\mathbb{C})((\lambda^{-1}))$, 
one can uniquely determine $H_{\ckp}^{[-1]}$ and $U_{\ckp}^{[-2]}$ from~\eqref{exam-bt-2}. %, as explained in the above proof.
\end{example}

An element $R^{\ckp}(\lambda)\in \mathcal{A}_{\ckp}[\epsilon]\otimes sl_{M+1}(\mathbb{C})((\lambda^{-1}))$ is called a {\it matrix resolvent (MR) of $\mathcal{L}_{\ckp}(\lambda)=\epsilon\partial_{X}+\Lambda^{\ckp}(\lambda)+V_{\ckp}$} if 
\begin{align}\label{resolv-def-eq-ckp}
		[\mathcal{L}_{\ckp}(\lambda),R^{\ckp}(\lambda)]=0.
\end{align}

The set of all matrix resolvents of $\mathcal{L}_{\ckp}(\lambda)$ is denoted by $\mathcal{M}_{\mathcal{L}_{\ckp}(\lambda)}$.
\begin{lemma}\label{kerlemma-ckp}
We have
\begin{align}
\mathcal{M}_{\mathcal{L}_{\ckp}(\lambda)}=e^{\ad_{U_{\ckp}} }\left(\mathfrak{h}_{\ckp}\right),
\end{align}
\end{lemma}
\begin{proof}
	By using Lemma~\ref{decomuni} and Lemma~\ref{funlemma}, the proof is similar to that in~\cite{BDY21}. So we omit its details.
\end{proof}

Following~\cite{BDY21}, we define the {\it extended gradation} by assigning the following degrees:
	\begin{align}
		&\deg^{e} e_{i,j}=j-i,\quad \deg^{e}\partial_{X}=1,\quad \deg^{e} \lambda=M, \quad i,j=1,\dots, M+1,\label{extend-deg-1}\\
		&\deg^{e} q=0,\quad \deg^{e} r=M+1, \quad \deg^{e} v_{l}=M-l,\quad l=0,\cdots, M-2.  \label{extend-deg-2}
	\end{align}
Like in~\cite{BDY21} we have $\deg^{e} \mathcal{L}_{\ckp}(\lambda)=1$.
We are ready to prove Lemma~\ref{lemmarnls}.
\begin{proof}[Proof of Lemma~\ref{lemmarnls}]
Let $U_{\ckp}$ be the element defined in Lemma \ref{funlemma}. 
Using Lemma \ref{kerlemma-ckp}, 
we know that $R_{a}^{\ckp}(\lambda)=e^{\ad_{U_{\ckp}}}(\Lambda_{a}^{\ckp}(\lambda)),1\le a \le M$, 
are solutions to equation~\eqref{reso-def-eq}.
Equation \eqref{ckpresolventcheng} follows from \eqref{U-form-ckp}, 
and equation \eqref{carconcheng} follows from the fact that
\begin{align}\label{leading-ckp}
	\tr(\Lambda_{a}^{\ckp}(\lambda)\Lambda_{b}^{\ckp}(\lambda))=\delta_{a+b,M}(a+b)\lambda+\delta_{a+b,2M}\frac{M}{M+1}\lambda^2.
\end{align}
This shows the existence of $R_{a}^{\ckp}(\lambda)$.
The uniqueness follows from Lemma \ref{kerlemma-ckp}. 
%The lemma is proved.	
\end{proof}
Recall that in Section~\ref{sec1}, 
we have defined a sequence of flows~\eqref{def-ckpflow} by using the basic matrix resolvents $R_{a}^{\ckp}(\lambda),1\le a\le M$. We first have the following lemma.
\begin{lemma}\label{uni-refine}
$\forall\, a=1,\dots,M, k\ge 0$, 	
there exists a unique correction term (see~\eqref{def-ckp-V-0612}) $\widehat{V}_{a,k}^{\ckp}\in\mathcal{A}_{\ckp}[\epsilon]\otimes sl_{M+1}(\mathbb{C})$,  
such that~\eqref{def-ckpflow} gives well-defined equations for $q,r, v_{0},\dots, v_{M-2}$. Moreover this unique correction term is given by~\eqref{core-ckp-v-0612}.
\end{lemma}
\begin{proof}
Write $	\widehat{V}_{a,k}^{\ckp}=\sum_{i=1}^{M+1}\sum_{j=1}^{M+1}\hat{v}_{a,k;i,j}^{\ckp}e_{i,j}, a=1,\dots,M, k\ge 0$.
By a direct calculation, we have
\begin{align*}
	&\left[V_{a,k}^{\ckp},\mathcal{L}_{\ckp}(\lambda)\right]\\
	=&\lambda\left(\sum_{i=1}^{M+1}\hat{v}_{a,k;i,M}^{\ckp}e_{i,1}-\hat{v}_{a,k;1,i}^{\ckp}e_{M,i}\right)+\sum_{i=1}^{M+1}\sum_{j=1}^{M+1}\epsilon\partial_{X}(\hat{v}_{a,k;i,j}^{\ckp})e_{i,j}\\
	&+\sum_{i=1}^{M-1}\sum_{j=2}^{M-1}(\hat{v}_{a,k;i,j-1}^{\ckp}-\hat{v}_{a,k;i+1,j}^{\ckp}+\hat{v}_{a,k;i,M}^{\ckp}v_{j-1})e_{i,j}                   \\
	&-\sum_{j=1}^{M+1}\left(r_{a,k+1;1,j}^{\ckp}+\sum_{i=1}^{M-1}\hat{v}_{a,k;i,j}^{\ckp}v_{i-1}-\hat{v}_{a,k;M+1,j}^{\ckp}q\right)e_{M,j}+\sum_{j=2}^{M-1}\left(\hat{v}_{a,k;M,M}^{\ckp}v_{j-1}-\hat{v}_{a,k;M,j}^{\ckp}\right)e_{M,j} \\
	&+\sum_{j=1}^{M+1}\hat{v}_{a,k;1,j}^{\ckp}re_{M+1,j}+\sum_{j=2}^{M-1}(\hat{v}_{a,k;M+1,j-1}^{\ckp}+\hat{v}_{a,k;M+1,M}^{\ckp}v_{j-1})e_{M+1,j}\\
	&+\sum_{i=1}^{M+1}(\hat{v}_{a,k;i,M}^{\ckp}v_{0}-\hat{v}_{a,k;i,M+1}^{\ckp})e_{i,1}-\sum_{i=1}^{M-1}\hat{v}_{a,k;i+1,1}^{\ckp}e_{i,1}+\sum_{i=1}^{M+1}\hat{v}_{a,k;i,M-1}^{\ckp}e_{i,M}-\sum_{i=1}^{M-1}\hat{v}_{a,k;i+1,M}^{\ckp}e_{i,M}.                                 
\end{align*}
The statements follow from this equation and~\eqref{def-ckpflow}. 
The lemma is proved.
\end{proof}
\begin{lemma}
The flows~\eqref{def-ckpflow} coincide with the constrained KP hierarchy~\eqref{laxeq-cheng}.
\end{lemma}
\begin{proof}
By using Lemma~\ref{lemmarnls}, we have
\begin{align*}
	V_{2,0}^{\ckp}=&\frac{2}{M}\sum_{l=0}^{M-1}\sum_{k=1}^{M-l}\binom{k+l-1}{l}v_{M-2,lX}e_{k+l,k}-\sum_{k=1}^{M-2}v_{k-1}e_{M-1,k}-\sum_{k=0}^{M-2}(v_{k-1}+v_{k,X})e_{M,k+1}\\
	&-v_{M-2}(e_{M-1,M-1}+e_{M,M})+qe_{M-1,M+1}+q_{X}e_{M,M+1}-re_{M+1,2}+r_{X}e_{M+1,1}\\
	&+\lambda(e_{M-1,1}+e_{M,2})+\sum_{k=1}^{M-2}e_{k,k+2}+\delta_{1,M}\left(\lambda^{2}e_{1,1}+\lambda(1+q)e_{2,1}-\lambda re_{2,1}+r(1-q)e_{2,2}\right),
\end{align*}
where $v_{-1}=qr$. Then, by using~\eqref{def-btflow}, we have
\begin{align*}
	\frac{\partial \mathcal{L}_{\ckp}}{\partial t^{2}_{0}}=&\epsilon^{-1}\sum_{k=0}^{M-2}\left(2\epsilon v_{k-1,X}+\epsilon^{2}v_{k,2X}-\frac{2}{M}\binom{M}{M-k}\epsilon^{M-k}v_{M-2,(M-k)X} \right.\\
	&\left. -\frac{2}{M}\sum_{i=k+1}^{M-2}\binom{i}{i-k}v_{i}\epsilon^{i-k}v_{M-2,(i-k)X}\right)e_{M,k+1}\\
	&-\left(\epsilon q_{2X}+\frac{2}{M}\epsilon^{-1}v_{M-2}q\right)e_{M,M+1}-\left(\epsilon r_{2X}+\frac{2}{M}\epsilon^{-1}v_{M-2}r\right)e_{M+1,1}.
\end{align*}
This implies 
\begin{align}
	&\frac{\partial v_{k}}{\partial t^{2}_{0}}=2v_{k-1,X}+\epsilon v_{k,2X}-\frac{2}{M}\binom{M}{M-k}\epsilon^{M-k-1}v_{M-2,(M-k)X}-\frac{2}{M}\sum_{i=k+1}^{M-2}\binom{i}{i-k}\epsilon^{i-k-1}v_{i}v_{M-2,(i-k)X},\\
	&\frac{\partial q}{\partial t^{2}_{0}}=\epsilon q_{2X}+\frac{2}{M}\epsilon^{-1}v_{M-2}q,\quad  \frac{\partial r}{\partial t^{2}_{0}}=-\epsilon r_{2X}-\frac{2}{M}\epsilon^{-1}v_{M-2}r,
\end{align}
where $k=0,\dots,M-2$.
This coincides with the $t^{2}_{0}$ flow of~\eqref{laxeq-bt}. 
Then by using the method similar to~\cite{BDY16} 
(see also the relevant calculations in~\cite{Y22, Y23} for the $M=1$ case), i.e., to match with the 
principal hierarchy of the corresponding Frobenius manifolds,  one can prove this lemma. 
\end{proof}

In the rest of this subsection, 
we will give some formulas which will be used in the  %for giving the tau-structure for the constrained KP hierarchy in the 
next subsection.
\begin{lemma}\label{rteq-ckp}
The following identities hold true
\begin{align}\label{rteq1}
		\epsilon\frac{\partial R_{b}^{\ckp}(\lambda)}{\partial t_{k}^{a}}=\left[V_{a,k}^{\ckp}(\lambda),R_{b}^{\ckp}(\lambda)\right], \quad 1\le a,b\le M,\,\, k\ge 0.
\end{align}
\end{lemma}
\begin{proof} 
	By using Lemma~\ref{funlemma} and~\eqref{def-ckpflow},  we have
\begin{align}
		\Big[\epsilon\p_{t_{k}^{a}} + S_{a,k}^{\ckp} , \epsilon\p_X + \Lambda^{\ckp}(\lambda)+H_{\ckp} \Big]=0,
\end{align}
where $S_{a,k}^{\ckp}=\sum_{l=0}^\infty \frac{(-1)^l}{(l+1)!} \ad_{U_{\ckp}}^{l} \bigl(\epsilon\frac{\p U_{\ckp}}{\p t_{k}^{a}}\bigr) - e^{-\ad_{U_{\ckp}}} \bigl(V_{a,k}^{\ckp}\bigr)\in \mathcal{A}_{\ckp}[\epsilon]\otimes sl_{M+1}(\mathbb{C})((\lambda^{-1}))$. 
Decompose 
\begin{align}
S_{a,k}^{\ckp}= S_{a,k}^{\ckp,\ke}+S_{a,k}^{\ckp,\im},
\end{align}
where $S_{a,k}^{\ckp,\ke} \in \mathcal{A}_{\ckp}[\epsilon]\otimes \ke\,\ad_{\Lambda^{\ckp}(\lambda)}, S_{a,k}^{\ckp,\im} \in \mathcal{A}_{\ckp}[\epsilon]\otimes \im\,\ad_{\Lambda^{\ckp}(\lambda)}$.
Then we have
\begin{align}
		& \epsilon\frac{\p H_{\ckp}}{\p t_{k}^{a} } -\epsilon \frac{\p S_{a,k}^{\ckp,\ke}}{\p X}=0, \\  
		& \epsilon \frac{\p S_{a,k}^{\ckp,\im}}{\p X}= [S_{a,k}^{\ckp,\im}, \Lambda^{\ckp}(\lambda)+H_{\ckp}].\label{imeq-ckp}
\end{align}
Using the argument similar to that in the proof of~\cite[Lemma 2.2.2]{BDY21} we find from~\eqref{imeq-ckp} that $S_{a,k}^{\ckp,\im}$ vanishes.
So $S_{a,k}^{\ckp}$ belongs to $\mathcal{A}_{\ckp}[\epsilon]\otimes \ke\,\ad_{\Lambda^{\ckp}(\lambda)}$.
Then we have
\begin{align*}
	0= \bigl[\epsilon\p_{t_{k}^{a}} + S_{a,k}^{\ckp} , \Lambda_{b}^{\ckp}(\lambda) \bigr]
	   =e^{-\ad_{U_{\ckp}}}\bigl[\epsilon \p_{t_{k}^{a}} - V_{a,k}^{\ckp}(\lambda),R_{b}^{\ckp}(\lambda) \bigr].
\end{align*}
The lemma is proved. 
\end{proof}

Define the loop operator by
\begin{align}
	\nabla_{a}(\lambda):=\sum_{k\ge 0}^{}\frac{1}{\lambda^{k+1}}\frac{\partial}{\partial t_{k}^{a}},\quad a=1,\dots,M.
\end{align}
\begin{lemma}\label{ckp-zero-eq}
We have the following identities
\begin{align}\label{imp-iden-ckp}
		\epsilon\nabla_{a}(\lambda)(R_{b}^{\ckp}(\mu))
		=\frac{\left[R_{a}^{\ckp}(\lambda),R_{b}^{\ckp}(\mu)\right]}{\lambda-\mu}
		+\left[Q_{a}^{\ckp}(\lambda),R_{b}^{\ckp}(\mu)\right],\quad 1\le a,b\le M,
\end{align}
where 
\begin{align}
	Q_{a}^{\ckp}(\lambda)=-\sum_{i=2}^{M}\sum_{j=1}^{i-1}\sum_{l=0}^{i-j-1}\binom{l+j-1}{j-1}(\epsilon\partial_{X})^{l}(r_{a;i-j-l,M}^{\ckp}(\lambda))e_{i,j}+\left(-\frac{1}{M+1}I+e_{M+1,M+1}\right)\delta_{a,M}.
\end{align}
\end{lemma}
\begin{proof}
By comparing the $\lambda$ coefficient of~\eqref{reso-def-eq} and by a direct calculation, we have
\begin{align}
	r_{a,0;i,M}^{\ckp}=\delta_{i,M-a},\quad 1\le i\le M-1,\quad 1\le a\le M.
\end{align}
Then,
\begin{align*}
	\sum_{k\ge 0}\frac{V_{a,k}^{\ckp}(\mu)}{\lambda^{k+1}}
	=&\frac{R_{a}^{\ckp}(\lambda)}{\lambda-\mu}-{\rm Coef}_{\lambda}(R_{a}^{\ckp}(\lambda))\\
	&-\sum_{i=2}^{M}\sum_{j=1}^{i-1}\sum_{l=0}^{i-j-1}\binom{l+j-1}{j-1}(\epsilon\partial_{X})^{l}\left(r_{a;i-j-l,M}^{\ckp}(\lambda)-r_{a,0;i-j-l,M}^{\ckp}-\lambda r_{a,-1;i-j-l,M}^{\ckp}\right)e_{i,j}\\
	=&\frac{R_{a}^{\ckp}(\lambda)}{\lambda-\mu}-\sum_{i=2}^{M}\sum_{j=1}^{i-1}\sum_{l=0}^{i-j-1}\binom{l+j-1}{j-1}(\epsilon\partial_{X})^{l}(r_{a;i-j-l,M}^{\ckp}(\lambda))e_{i,j}+\left(-\frac{1}{M+1}I+e_{M+1,M+1}\right)\delta_{a,M}.
\end{align*}
The lemma is proved by using~\eqref{rteq1}.
\end{proof}
\begin{lemma}\label{zero-ckp}
 We have the following identities
\begin{align}\label{rteq2}
	       \epsilon\frac{\partial V_{a,i}^{\ckp}(\lambda)}{\partial t_{j}^{b}}-\epsilon\frac{\partial V_{b,j}^{\ckp}(\lambda)}{\partial t_{i}^{a}}+\left[V_{a,i}^{\ckp}(\lambda),V_{b,j}^{\ckp}(\lambda)\right]=0,\quad 1\le a,b\le M,\quad i,j\ge 0.
\end{align}
\end{lemma}
\begin{proof}
Using \eqref{imp-iden-ckp}, we obtain
\begin{align*}
	       &\sum_{i,j\ge 0}\frac{1}{\lambda^{i+1}\mu^{j+1}}\left( \epsilon\frac{\partial V_{a,i}^{\ckp}(\nu)}{\partial t_{j}^{b}}-\epsilon\frac{\partial V_{b,j}^{\ckp}(\nu)}{\partial t_{i}^{a}}+\left[V_{a,i}^{\ckp}(\nu),V_{b,j}^{\ckp}(\nu)\right]\right)\\
%		=&\epsilon\nabla_{b}(\mu)(V_{a}^{\ckp}(\lambda,\nu))-\epsilon\nabla_{a}(\lambda)(V_{b}^{\ckp}(\mu,\nu))+[V_{a}^{\ckp}(\lambda,\nu),V_{b}^{\ckp}(\mu,\nu)]\\
%		&+\left[\frac{R_{a}^{\ckp}(\lambda)}{\lambda-\mu}+Q_{a}^{\ckp}(\lambda),\frac{R_{b}^{\ckp}(\mu)}{\mu-\nu}+Q_{b}^{\ckp}(\mu)\right]\\
		=&\frac{[R_{b}^{\ckp}(\mu),R_{a}^{\ckp}(\lambda)]}{(\lambda-\nu)(\mu-\lambda)}+\frac{[Q_{b}^{\ckp}(\mu),R_{a}^{\ckp}(\lambda)]}{\lambda-\nu}+\epsilon\nabla_{b}(\mu)(Q_{a}^{\ckp}(\lambda))\\
		&-\frac{[R_{a}^{\ckp}(\lambda),R_{b}^{\ckp}(\mu)]}{(\mu-\nu)(\lambda-\mu)}-\frac{[Q_{a}^{\ckp}(\lambda),R_{b}^{\ckp}(\mu)]}{\mu-\nu}-\epsilon\nabla_{a}(\lambda)(Q_{b}^{\ckp}(\mu))\\
		&+\left[\frac{R_{a}^{\ckp}(\lambda)}{\lambda-\mu},\frac{R_{b}^{\ckp}(\mu)}{\mu-\nu}\right]+\left[\frac{R_{a}^{\ckp}(\lambda)}{\lambda-\mu},Q_{b}^{\ckp}(\mu)\right]+\left[Q_{a}^{\ckp}(\lambda),\frac{R_{b}^{\ckp}(\mu)}{\mu-\nu}\right]+\left[Q_{a}^{\ckp}(\lambda),Q_{b}^{\ckp}(\mu)\right]\\
		=&\epsilon\nabla_{b}(\mu)(Q_{a}^{\ckp}(\lambda))-\epsilon\nabla_{a}(\lambda)(Q_{b}^{\ckp}(\mu))+\left[Q_{a}^{\ckp}(\lambda),Q_{b}^{\ckp}(\mu)\right].
\end{align*}
This means that the validity of~\eqref{rteq2} is equivalent to
\begin{align}
	\epsilon\nabla_{b}(\mu)(Q_{a}^{\ckp}(\lambda))-\epsilon\nabla_{a}(\lambda)(Q_{b}^{\ckp}(\mu))+\left[Q_{a}^{\ckp}(\lambda),Q_{b}^{\ckp}(\mu)\right]=0.
\end{align}
By using~\eqref{imp-iden-ckp} and by a direct calculation, we have 
\begin{align*}
	\epsilon\sum_{i=1}^{M+1}\nabla_{a}(\lambda)(r_{b;i,M}^{\ckp}(\mu))e_{i,M}=&\frac{1}{\lambda-\mu}\left(\sum_{i=1}^{M+1}\sum_{j=1}^{M+1}r_{a;i,j}^{\ckp}(\lambda)r_{b;j,M}^{\ckp}(\mu)-\sum_{i=1}^{M+1}\sum_{j=1}^{M+1}r_{b;i,j}^{\ckp}(\mu)r_{a;j,M}^{\ckp}(\lambda)\right)\\
	&-\sum_{i=2}^{M}\sum_{k=1}^{i-1}\left[Q_{a}^{\ckp}(\lambda)\right]_{i,k}r_{b;k,M}^{\ckp}(\mu)e_{i,M}+r_{b;M+1,M}^{\ckp}(\mu)e_{M+1,M}\delta_{a,M}.
\end{align*}
Also we have
\begin{align*}
	\epsilon\sum_{i=1}^{M+1}\nabla_{b}(\mu)(r_{a;i,M}^{\ckp}(\lambda))e_{i,M}=&\frac{1}{\mu-\lambda}\left(\sum_{i=1}^{M+1}\sum_{j=1}^{M+1}r_{b;i,j}^{\ckp}(\mu)r_{a;j,M}^{\ckp}(\lambda)-\sum_{i=1}^{M+1}\sum_{j=1}^{M+1}r_{a;i,j}^{\ckp}(\lambda)r_{b;j,M}^{\ckp}(\mu)\right)\\
	&-\sum_{i=2}^{M}\sum_{k=1}^{i-1}\left[Q_{b}^{\ckp}(\mu)\right]_{i,k}r_{a;k,M}^{\ckp}(\lambda)e_{i,M}+r_{a;M+1,M}^{\ckp}(\lambda)e_{M+1,M}\delta_{b,M}.
\end{align*}
Then, by a direct calculation, we have
\begin{align*}
	&\epsilon\nabla_{a}(\lambda)(Q_{b}^{\ckp}(\mu))-\epsilon\nabla_{b}(\mu)(Q_{a}^{\ckp}(\lambda))+\left[Q_{b}^{\ckp}(\mu),Q_{a}^{\ckp}(\lambda)\right]\\
	=&\sum_{j=1}^{M-1}\sum_{i=j+1}^{M}\sum_{k=1}^{i-j-1}\sum_{\alpha=0}^{k-1}\sum_{\beta=0}^{i-j-k-1}\left(\binom{\beta+j-1}{j-1}\binom{\alpha+k+j-1}{k+j-1}\right.\\
	&\qquad\qquad\qquad\qquad\qquad\qquad\qquad\left.-\sum_{l=0}^{\alpha}\binom{l+\beta+j-1}{j-1}\binom{l+\beta}{l}\binom{\alpha-l+k-\beta-1}{k-\beta-1}\right)\\
	&\cdot\left((\epsilon\partial_{X})^{\alpha}(r_{a;k-\alpha,M}^{\ckp}(\lambda))(\epsilon\partial_{X})^{\beta}(r_{b;i-j-k-\beta,M}^{\ckp}(\mu))-(\epsilon\partial_{X})^{\alpha}(r_{b;k-\alpha,M}^{\ckp}(\mu))(\epsilon\partial_{X})^{\beta}(r_{a;i-j-k-\beta,M}^{\ckp}(\lambda))\right)e_{i,j}.
\end{align*}
Note that 
\begin{align*}
	&\sum_{l=0}^{\alpha}\binom{l+\beta+j-1}{j-1}\binom{l+\beta}{l}\binom{\alpha-l+k-\beta-1}{k-\beta-1}\\
	=&\binom{\beta+j-1}{j-1}\sum_{l=0}^{\alpha}\binom{l+\beta+j-1}{l}\binom{\alpha-l+k-\beta-1}{\alpha-l}\\
%	=&\binom{\beta+j-1}{j-1}\sum_{l=0}^{\alpha}(-1)^{l}\binom{-\beta-j}{l}(-1)^{\alpha-l}\binom{-k+\beta}{\alpha-l}\\
	\overset{\star}{=}&\binom{\beta+j-1}{j-1}(-1)^{\alpha}\binom{-k-j}{\alpha}=\binom{\beta+j-1}{j-1}\binom{\alpha+j+k-1}{\alpha},
\end{align*}
where in the step ``$\star$" we used the Chu--Vandermonde identity~\cite{A75}.
The lemma is proved.
\end{proof}
\begin{lemma}\label{comu-ckp-pro}
	The flows~\eqref{def-ckpflow} pairwise commute.
\end{lemma}
\begin{proof}
Follows from~\eqref{zero-ckp} and the definition of these flows.
\end{proof}

\subsection{From matrix resolvents to tau-functions for the constrained KP hierarchy }\label{subsec2.2}
Let us first show that $\Omega_{a,i;b,j}^{\ckp}$ are well defined by~\eqref{def-ckp-tau-str}. 
Indeed,
\begin{align}
		R_{b}^{\ckp}(\mu) = R_{b}^{\ckp}(\lambda)+ \bigl(R_{b}^{\ckp}(\lambda)\bigr)^{'} (\mu-\lambda) + 
		(\mu-\lambda)^2 \, \p_\lambda \left(\frac{R_{b}^{\ckp}(\lambda)-R_{b}^{\ckp}(\mu)}{\lambda-\mu}\right)
\end{align}
and using~\eqref{leading-ckp} we have
\begin{align}
		\frac{\tr \left(R_{a}^{\ckp} (\lambda) R_{b}^{\ckp}(\mu)\right)}{(\lambda-\mu)^2} =& \frac{\delta_{a+b,M}(a+b)\lambda+\delta_{a+b,2M}\frac{M}{M+1}\lambda^2}{(\lambda-\mu)^{2}}- \frac{\tr \bigl(R_{a}^{\ckp}(\lambda) \bigl(R_{b}^{\ckp}(\lambda)\bigr)^{'}\bigr)}{\lambda-\mu}   \nonumber\\
		& + \tr \biggl(R_{a}^{\ckp}(\lambda)\, \p_\lambda \left(\frac{R_{b}^{\ckp}(\lambda)-R_{b}^{\ckp}(\mu)}{\lambda-\mu}\right)\biggr),
\end{align}
where $``\,'\,"$ denotes derivative with respect to the spectral parameter.
Since $R_{a}^{\ckp}(\lambda) = \mathcal{O}(\lambda^1), a=1,\dots,M$, it can be deduced that the third term in the identity above has the same form as the left-hand side of \eqref{def-ckp-tau-str}. Thus, the only thing left to prove is that
\begin{align}
	\frac{\delta_{a+b,M}M\lambda+\delta_{a+b,2M}\frac{M}{M+1}\lambda^2}{(\lambda-\mu)^{2}} 
	- \frac{\tr \bigl(R_{a}^{\ckp}(\lambda) \bigl(R_{b}^{\ckp}(\lambda)\bigr)^{'}\bigr)}{\lambda-\mu} -\frac{a\lambda+b\mu}{(\lambda-\mu)^{2}}\delta_{a+b,M}-\frac{M}{M+1}\frac{\lambda \mu}{(\lambda-\mu)^{2}}\delta_{a+b,2M}
\end{align}
has the same form as the left-hand side of \eqref{def-ckp-tau-str}. It will be shown that the above expression equals zero.
Indeed,
\begin{align*}\label{x-ind}
		&\partial_{X} \left(\tr \left(R_{a}^{\ckp}(\lambda)  \bigl(R_{b}^{\ckp}(\lambda)\bigr)^{'}\right) \right)  \nonumber\\
		= &\tr \left(\left[R_{a}^{\ckp}(\lambda),\Lambda^{\ckp}(\lambda)+V_{\ckp}\right] \bigl(R_{b}^{\ckp}(\lambda)\bigr)^{'}\right)+\tr \left(R_{a}^{\ckp}(\lambda)\left( \left[\bigl(R_{b}^{\ckp}(\lambda)\bigr),\Lambda^{\ckp}(\lambda)+V_{\ckp}\right]\right)^{'}\right) \nonumber\\
			%+\left[R_{b}^{\ckp}(\lambda),(\Lambda^{\ckp}(\lambda)\bigr)^{'}\right]\right)\right)\nonumber \\
			=&\tr \left(R_{a}^{\ckp} (\lambda)\left[R_{b}^{\ckp}(\lambda),(\Lambda^{\ckp}(\lambda)\bigr)^{'}\right]\right)
			=\tr \left(\left[R_{b}^{\ckp}(\lambda),R_{a}^{\ckp}(\lambda)\bigl(\Lambda^{\ckp}(\lambda)\bigr)^{'}\right]\right)=0.
\end{align*}
Here we have used the commutativity between matrix resolvents.
Therefore, we conclude that $\tr \left(R_{a}^{\ckp}(\lambda) \bigl(R_{b}^{\ckp}(\lambda)\bigr)^{'}\right)$ is independent of $q,r,v_{0},\dots,v_{M-2}$ and all orders of $X$-derivative, meaning that it is only a function of $\lambda$. Consequently, we obtain
\begin{align}
	\tr \left(R_{a}^{\ckp}(\lambda) \bigl(R_{b}^{\ckp}(\lambda)\bigr)^{'}\right)= \tr \left(\Lambda_{a}^{\ckp}(\lambda) \bigl(\Lambda_{b}^{\ckp}(\lambda)\bigr)^{'}\right)=b\delta_{a+b,M}+\delta_{a+b,2M}\frac{M}{M+1}\lambda.
\end{align}
Therefore, $\Omega_{a,i;b,j}^{\ckp}$ are well defined.

%Let us now prove Lemma~\ref{tau-str-ckp}
\begin{proof}[Proof of Lemma~\ref{tau-str-ckp}]
By using~\eqref{imp-iden-ckp}, the proof is similar to that in~\cite{BDY21,DY17}. We omit its details.
\end{proof}
\begin{proof}[Proof of Proposition~\ref{main-thm} ]
The proof is similar to that in~\cite{BDY21,DY17}, so we omit its details.
\end{proof}
Write $R_{a}^{\ckp}({\lambda;{\bf t}};\epsilon)$ as the $R_{a}^{\ckp}(\lambda)$ evaluated at an arbitrarily solution~\eqref{solution-ckp-cheng-sec1}. The following corollary follows from Proposition~\ref{main-thm}.
\begin{cor}
For any integer $k\ge 2$, and any fixed $a_{1},\dots,a_{k}\in \{1,\dots,M\}$, we have
\begin{align}
		\sum_{i_{1},\dots,i_{k}\ge 0}\frac{\epsilon^{k}\frac{\partial^{k} \log \tau_{\ckp}({\bf t};\epsilon)}{\partial t^{a_{1}}_{i_{1}}\cdots \partial t^{a_{k}}_{i_{k}}}}{\prod_{j=1}^{k}\lambda_{j}^{i_{k}+1}}
		=&-\sum_{\sigma\in S_{k}/C_{k}}\frac{\tr \left(R_{a_{\sigma(1)}}^{\ckp}(\lambda_{\sigma(1)};{\bf t};\epsilon)R_{a_{\sigma(2)}}^{\ckp}(\lambda_{\sigma(2)};{\bf t};\epsilon)\cdots R_{a_{\sigma(k)}}^{\ckp}(\lambda_{\sigma(k)};{\bf t};\epsilon)\right)}{\prod_{i=1}^{k}(\lambda_{\sigma(i)}-\lambda_{\sigma(i+1)})} \nonumber\\
		&-\left(\frac{a_{1}\lambda_{1}+a_{2}\lambda_{2}}{(\lambda_{1}-\lambda_{2})^{2}}\delta_{a_{1}+a_{2},M}+\frac{M}{M+1}\frac{\lambda_{1} \lambda_{2}}{(\lambda_{1}-\lambda_{2})^{2}}\delta_{a_{1}+a_{2},2M}\right)\delta_{k,2}.
\end{align}
\end{cor}

\section{The matrix-resolvent method to tau-functions for the  bigraded Toda  hierarchy of $(M,1)$-type}	\label{sec3}

\subsection{Matrix resolvents for the  bigraded Toda hierarchy of $(M,1)$-type}
For any $A\in \mathcal{A}[[\epsilon]]\otimes \mat(M+1,\mathbb{C}((\lambda^{-1})))$, we define
\begin{align}
	\rho_{A}: \mathcal{A}[[\epsilon]]\otimes \mat(M+1,\mathbb{C}((\lambda^{-1})))\to \mathcal{A}[[\epsilon]]\otimes \mat(M+1,\mathbb{C}((\lambda^{-1}))) 
\end{align}
by
\begin{align}
	\rho_{A}(B):=AB-\mathcal{T}(B)A,\quad \forall B\in\mathcal{A}[[\epsilon]]\otimes \mat(M+1,\mathbb{C}((\lambda^{-1}))).
\end{align}
Let us denote $\tilde{\rho}_{A}:=\rho_{A}|_{\mat(M+1,\mathbb{C}((\lambda^{-1})))}$. We have the following lemma.
\begin{lemma}\label{ker-im-bt}
The spaces $\im\tilde{\rho}_{\Lambda(\lambda)},\ke\tilde{\rho}_{\Lambda(\lambda)}\subset \mat(M+1,\mathbb{C}((\lambda^{-1})))$ have the following form	
\begin{align}\label{bt-decom}
	&\im\tilde{\rho}_{\Lambda(\lambda)}=\left\{\sum_{l=1}^{M^{2}+M}\alpha_{l}(\lambda)e_{l}(\lambda)\bigg|\alpha_{l}(\lambda)\in \mathbb{C}((\lambda^{-1}))\right\},\\
	&\ke\tilde{\rho}_{\Lambda(\lambda)}=\left\{\beta_{0}(\lambda)I+\sum_{a=1}^{M}\beta_{a}(\lambda)\Lambda_{a}(\lambda)\bigg|\beta_{0}(\lambda),\beta_{a}(\lambda)\in\mathbb{C}((\lambda^{-1}))\right\}=:\mathfrak{h}.
\end{align}
Here
\begin{align*}
	&e_{k(M-1)+i}(\lambda)=\left\{\begin{aligned}
		&	\lambda e_{1,k+1}-\lambda e_{i+1,i+k+1}\quad  & i=1,\dots ,M-k-1\\
		&	\lambda e_{1,k+1}-e_{i+1,i+k+1-M}\quad       & i=M-k,\dots, M-1
	\end{aligned}  
	\right.,   \quad k=1,\dots, M-1,\\
	&e_{M(M-1)+i}(\lambda)=\lambda e_{i+1,M+1}-e_{i+1,1},\quad i=1,\dots ,M-1,\\
	&e_{(M+1)(M-1)+1}(\lambda)=\lambda e_{1,M+1}+e_{M+1,M+1}-e_{1,1},\quad e_{(M+1)(M-1)+a+1}(\lambda)=e_{M+1,a},\quad a=1,\dots, M.
\end{align*}
\end{lemma}
\noindent The proof of this lemma is in Appendix~\ref{appen-A}.  

Note that $e_{l}(\lambda),\Lambda_{a}(\lambda),I,1\le l\le M^{2}+M, 1\le a\le M$, are linearly independent. We have
\begin{align}\label{deco-bt}
	\mat(M+1,\mathbb{C}((\lambda^{-1})))=\im\tilde{\rho}_{\Lambda(\lambda)} \oplus\ke\tilde{\rho}_{\Lambda(\lambda)}.
\end{align}
It is easy to check that for any $A\in \mathcal{A}[[\epsilon]]\otimes \im \tilde{\rho}_{\Lambda(\lambda)}$, 
 $\rho_{\Lambda(\lambda)}(A)\in \mathcal{A}[[\epsilon]]\otimes \im \tilde{\rho}_{\Lambda(\lambda)}$. 
\begin{lemma}\label{uniquelemma}
For any $B\in \mathcal{A}[[\epsilon]]\otimes\im\tilde{\rho}_{\Lambda(\lambda)}$, there exists a unique $A\in\mathcal{A}[[\epsilon]]\otimes \im\tilde{\rho}_{\Lambda(\lambda)}$ such that
\begin{align}\label{inv-bt}
		\rho_{\Lambda(\lambda)}(A)=B.
\end{align}
\end{lemma}
\begin{proof}
Write 
\begin{align}
		A=\sum_{l=1}^{M^{2}+M}a_{l}e_{l}(\lambda),\quad B=\sum_{l=1}^{M^{2}+M}b_{l}e_{l}(\lambda),\quad a_{l},b_{l}\in\mathcal{A}[[\epsilon]]\otimes \mathbb{C}((\lambda^{-1})).
\end{align}
By a direct calculation, we have
\begin{align*}
		&	\Lambda(\lambda)e_{i}(\lambda)=\left\{ \begin{aligned}
			&	  \lambda e_{(M-1)^{2}+1}-\lambda e_{(M-1)^{2}+i+1} \quad i=1,\dots,M-2\\
			&	  \lambda e_{(M-1)^{2}+1}+\lambda e_{(M+1)M}  \quad i=M-1
		\end{aligned} \right.,\\
		&\Lambda(\lambda)e_{k(M-1)+i}(\lambda)=\left\{ \begin{aligned}
			&	   e_{(k-1)(M-1)+1}- e_{(k-1)(M-1)+i+1}  \quad i=1,\dots,M-2 \\
			&	   e_{(k-1)(M-1)+1}+ e_{(M+1)M-k}  \quad i=M-1 
		\end{aligned} \right., \quad k=1,\dots, M-1,\\
		& \Lambda(\lambda) e_{M(M-1)+i}(\lambda)=\left\{\begin{aligned}
			-e_{M(M-2)+i+1}  \quad i=1,\dots,M-2 \\
			-\lambda e_{M(M-1)+M}+e_{M(M-1)+M+1} \quad i=M-1 
		\end{aligned} \right.,\\
		&\Lambda(\lambda)e_{(M+1)(M-1)+1}(\lambda)=-e_{M(M-1)+1}+\delta_{M+1,2}\left(-\lambda e_{1}+e_{1}+e_{2}\right),\\
		&\Lambda(\lambda) e_{(M+1)(M-1)+a+1}(\lambda)=0,\quad a=1,\dots, M,\\
		& e_{i}(\lambda)\Lambda(\lambda)=-\lambda e_{(M-1)^{2}+i},\quad e_{M(M-1)+i}(\lambda)\Lambda(\lambda)=0,\quad i=1,\dots, M-1,\\
		& e_{k(M-1)+i}(\lambda)\Lambda(\lambda)=-e_{(k-1)(M-1)+i},\quad i=1,\dots,M-1, k=1,\dots,M-1,\\
		& e_{(M+1)(M-1)+1}(\lambda)\Lambda(\lambda)=-e_{(M+1)M},\quad e_{(M+1)(M-1)+2}(\lambda)\Lambda(\lambda)=-\lambda e_{(M+1)M},\\
		& e_{(M+1)(M-1)+i+2}(\lambda)\Lambda(\lambda)=-e_{(M+1)(M-1)+i+1},\quad i=1,\dots,M-1.
\end{align*}
Then equation \eqref{inv-bt} is equivalent to
\begin{align*}
		&	\lambda \sum_{i=1}^{M-1}a_{i}+\lambda \mathcal{T}(a_{1})=b_{(M-1)^{2}+1},\\
		&   -\lambda a_{l}+\lambda \mathcal{T}(a_{l+1})=b_{(M-1)^{2}+l+1},\quad l=1,\dots,M-2,\\
		&	 \sum_{i=1}^{M-1}a_{k(M-1)+i}+ \mathcal{T}(a_{k(M-1)+1})=b_{k(M-1)+1},\quad k=1,\dots,M-1,\\
		&   - a_{k(M-1)+l}+ \mathcal{T}(a_{k(M-1)+l+1})=b_{k(M-1)+l+1},\quad l=1,\dots,M-2,\\	
		&   -a_{M(M-1)+l}=b_{M(M-1)+l+1},\quad l=1,\dots, M-2,\\
		&  -\lambda a_{M(M-1)+(M-1)}=b_{(M+1)(M-1)+1},\\
		&  -a_{(M+1)(M-1)+1}+\delta_{M+1,2}(-\lambda a_{1}+a_{1})=b_{M(M-1)+1},\\
		&  \lambda a_{M-1}+\mathcal{T}(a_{(M+1)(M-1)+1})+\lambda \mathcal{T}(a_{(M+1)(M-1)+2})+\delta_{M+1,2}a_{1}=b_{(M+1)M},\\
		&  a_{(i+1)(M-1)+M-1}+\mathcal{T}(a_{(M+1)(M-1)+i+2})=b_{(M+1)(M-1)+i+1},\quad i=1,\dots,M-1.
\end{align*}
By solving these equations we obtain
\begin{align*}
		&a_{k(M-1)+1}=\frac{1}{\lambda}(1+\mathcal{T}+\mathcal{T}^{2}+\dots+\mathcal{T}^{M-1})^{-1}\left(\mathcal{T}^{M-2}(b_{k(M-1)+1})-\sum_{i=1}^{M-1}\sum_{q=1}^{i-1}\mathcal{T}^{M-2+q-i}(b_{k(M-1)+q+1})\right),\\
		&a_{k(M-1)+i}=\frac{1}{\lambda}\sum_{q=1}^{i-1}\mathcal{T}^{-i+q}(b_{k(M-1)+q+1})+\mathcal{T}^{-i+1}(a_{k(M-1)+1}),\quad i=2,\dots,M-1,\\
		&   a_{M(M-1)+l}=-b_{M(M-1)+l+1},\quad l=1,\dots, M-2,\\
		&  a_{M(M-1)+(M-1)}=-\frac{1}{\lambda}b_{(M+1)(M-1)+1},\\
		&  a_{(M+1)(M-1)+1}=\delta_{M+1,2}(-\lambda a_{1}+a_{1})-b_{M(M-1)+1},\\
		&  a_{(M+1)(M-1)+2}=\frac{1}{\lambda}(\mathcal{T}^{-1}(b_{(M+1)M})-\delta_{M+1,2}\mathcal{T}^{-1}(a_{1})-\lambda \mathcal{T}^{-1}(a_{M-1})-a_{(M+1)(M-1)+1}),\\
		&  a_{(M+1)(M-1)+i+2}=\mathcal{T}^{-1}(b_{(M+1)(M-1)+i+1}-a_{(i+1)(M-1)+M-1}),\quad i=1,\dots,M-1.
\end{align*}
The lemma is proved.
\end{proof}

We have the following fundamental lemma.
\begin{lemma}\label{fundmenlemma}
Let $\mathcal{L}(\lambda)=\mathcal{T}+\Lambda(\lambda)+V$.
There exists a unique pair $(U,H)$ of the form
\begin{align}
		&	U=\sum_{k\ge 1}U^{[-k]}\in\mathcal{A}[[\epsilon]]\otimes \im \tilde{\rho}_{\Lambda(\lambda)},\label{U-form}\\
		&	 H=\sum_{k\ge 0}H^{[-k]}\in\mathcal{A}[[\epsilon]]\otimes \ke \tilde{\rho}_{\Lambda(\lambda)}, \label{H-form}
\end{align}
where $\im ,\ke $ are taken in $\mat(M+1,\mathbb{C}((\lambda^{-1})))$, such that
\begin{align}\label{fundmental-eq}
		e^{-\mathcal{T}(U)}\circ\mathcal{L}(\lambda)\circ e^{U}=\mathcal{T}+\Lambda(\lambda)+H,
\end{align}
where $U^{[-k]},H^{[-k]}$ denote the element in $\mathcal{A}[[\epsilon]]\otimes \mat(M+1,\mathbb{C}((\lambda^{-1})))$ with degree $-k$.
\end{lemma}
\begin{proof} Note that equation \eqref{fundmental-eq} is equivalent to
\begin{align}\label{recurrUHbt}
		\Biggl(\sum_{k\ge 0}\frac{(-1)^{k}\left(\mathcal{T}\left(U\right)\right)^{k}}{k!}\Biggr)\left(\Lambda(\lambda)+V\right)\Biggl(\sum_{k\ge 0}\frac{U^{k}}{k!}\Biggr)=\Lambda(\lambda)+H.
\end{align}
Substituting~\eqref{U-form} and~\eqref{H-form} into~\eqref{recurrUHbt} and 
comparing the components of both sides of~\eqref{recurrUHbt}  with degree $-k, k\ge 0$, we get
\begin{align}\label{recurbt}
   \rho_{\Lambda(\lambda)}\left(U^{[-k-1]}\right)+G_k=H^{[-k]}, 
\end{align}
where $G_k\in \mathcal{A}[[\epsilon]]\otimes\mat(M+1,\mathbb{C}((\lambda^{-1}))$, which are determined by $V,U^{[-1]},\dots,U^{[-k]}$.
We will prove the lemma by induction on the degree.
First, for $k=0$ equation \eqref{recurbt} reads
\begin{align}\label{recur0-bt}
	\rho_{\Lambda(\lambda)}\left(U^{[-1]}\right)+V^{[0]}=H^{[0]}.
\end{align}
Decompose
\begin{align}
V^{[0]}=V^{[0],\im}+V^{[0],\ke},
\end{align}
where $V^{[0],\im}\in \mathcal{A}[[\epsilon]]\otimes \im \tilde{\rho}_{\Lambda(\lambda)}, V^{[0],\ke}\in \mathcal{A}[[\epsilon]]\otimes \ke \tilde{\rho}_{\Lambda(\lambda)}$.
Then by using~\eqref{deco-bt}, 
equation~\eqref{recur0-bt} is equivalent to
\begin{align}
	H^{[0]}=V^{[0],\ke},\quad \rho_{\Lambda(\lambda)}\left(U^{[-1]}\right)=-V^{[0],\im}.
\end{align}
Using Lemma~\ref{uniquelemma}, 
we have $U^{[-1]}=\rho_{\Lambda(\lambda)}^{-1}\left(-V^{[0],\im}\right)$.
The second step of the induction follows from equations \eqref{recurbt},~\eqref{deco-bt} and Lemma~\ref{uniquelemma}. 
The lemma is proved.
\end{proof}
\begin{example}
Comparing the degree $-1$ parts of both sides of equation~\eqref{recurrUHbt},
we obtain
\begin{align}
		\frac{1}{2}\rho_{\rho_{\Lambda(\lambda)}(U^{[-1]})}(U^{[-1]})+\rho_{V^{[0]}}(U^{[-1]})+V^{[-1]}=H^{[-1]}-\rho_{\Lambda(\lambda)}(U^{[-2]}).
\end{align}
Since $U^{[-2]}\in \mathcal{A}[[\epsilon]]\otimes \im \tilde{\rho}_{\Lambda(\lambda)},H^{[-1]}\in \mathcal{A}[[\epsilon]]\otimes \ke \tilde{\rho}_{\Lambda(\lambda)}$, and the left-hand side belongs to $\mathcal{A}^{\ckp}[\epsilon]\otimes \mat(M+1,\mathbb{C}((\lambda^{-1})))$, 
we can uniquely determine $H^{[-1]}$ and $U^{[-2]}$, as explained in the above proof.
\end{example}

An element $R(\lambda)\in \mathcal{A}[[\epsilon]]\otimes \mat(M+1,\mathbb{C}((\lambda^{-1})))$ is called a {\it MR of $\mathcal{L}(\lambda)=\mathcal{T}+\Lambda(\lambda)+V$} if 
\begin{align}\label{resolv-def-eq}
		\mathcal{T}(R(\lambda))(\Lambda(\lambda)+V)-(\Lambda(\lambda)+V)R(\lambda)=0.
\end{align}

The set of all matrix resolvents of $\mathcal{L}(\lambda)$ is denoted by $\mathcal{M}_{\mathcal{L}(\lambda)}$.
\begin{lemma}\label{kerlemma}
We have
\begin{align}
		\mathcal{M}_{\mathcal{L}(\lambda)}=e^{U }\circ \mathfrak{h} \circ e^{-U}.
\end{align}
\end{lemma}
\begin{proof}
Lemma~\ref{fundmenlemma} reduces the problem to considering the resolvents of $\mathcal{T}+\Lambda(\lambda)+H$.
So, let us look at the following equation for $R_{H}(\lambda)\in \mathcal{A}[[\epsilon]]\otimes \mat(M+1,\mathbb{C}((\lambda^{-1})))$:
%Let $(U,H)$ be the pair defined in Lemma~\ref{fundmenlemma}, 
%then~\eqref{resolv-def-eq} is equivalent to
\begin{align}\label{bt-rec-eq}
 \rho_{\Lambda(\lambda)+H}\bigl(R_{H}(\lambda)\bigr)=0.
%	\mathcal{T}(R_{H}(\lambda))(\Lambda(\lambda)+H)-(\Lambda(\lambda)+H)R_{H}(\lambda)=0.
\end{align}	
%where $R_{H}(\lambda):=e^{-U}\circ R(\lambda)\circ e^{U}\in \mathcal{A}[[\epsilon]]\otimes \mat(M+1,\mathbb{C}((\lambda^{-1})))$.
By using~\eqref{deco-bt}, we write
\begin{align}
&	R_{H}(\lambda)=R_{H}^{\ke}(\lambda)+R_{H}^{\im}(\lambda),
\end{align}
where $R_{H}^{\ke}(\lambda)\in \mathcal{A}[[\epsilon]]\otimes \ke \tilde{\rho}_{\Lambda(\lambda)},R_{H}^{\im}(\lambda)\in\mathcal{A}[[\epsilon]]\otimes \im \tilde{\rho}_{\Lambda(\lambda)}$.
Then equation~\eqref{bt-rec-eq} holds if and only if
\begin{align}
\rho_{\Lambda(\lambda)+H}\left(R_{H}^{\im}(\lambda)\right)+\rho_{\Lambda(\lambda)+H}\left(R_{H}^{\ke}(\lambda)\right)=0.
\end{align}
It is easy to check that $\rho_{\Lambda(\lambda)+H}\left(R_{H}^{\im}(\lambda)\right)\in \mathcal{A}[[\epsilon]]\otimes \im \tilde{\rho}_{\Lambda(\lambda)},\rho_{\Lambda(\lambda)+H}\left(R_{H}^{\ke}(\lambda)\right)\in \mathcal{A}[[\epsilon]]\otimes \ke \tilde{\rho}_{\Lambda(\lambda)}$. 
Then, by using~\eqref{deco-bt}, we have 
\begin{align}\label{uni1}
	&\rho_{\Lambda(\lambda)+H}\left(R_{H}^{\ke}(\lambda)\right)=0,\quad  \rho_{\Lambda(\lambda)+H}\left(R_{H}^{\im}(\lambda)\right)=0.
\end{align}
The first equation of \eqref{uni1} is equivalent to 
%equation \eqref{uni1} is equivalent to 
\begin{align}\label{uni1-bt}
\mathcal{T}\bigl(R_{H}^{\ke}(\lambda)\bigr)-R_{H}^{\ke}(\lambda)=0.
%	(\mathcal{T}-1)(R_{H}^{\ke}(\lambda))=0.
\end{align}
This implies that $R_{H}^{\ke}(\lambda)$ depends only on $\lambda$. 
Let us now show that $R_{H}^{\im }(\lambda)$ must vanish. If it does not vanish, then there exists an integer $d$ such that 
$$
R_{H}^{\im}(\lambda)= \sum_{i=-\infty}^d R_{H}^{[i],\im}(\lambda),\qquad R_{H}^{[d],\im}(\lambda)\neq 0,
$$ 
where $R_{H}^{[i],\im}(\lambda)$ denotes the degree $i$ term of $R_{H}^{\im}(\lambda)$ (see~\eqref{bt-degree}).
Looking at the highest degree term on both sides of the second equation of~\eqref{uni1} we obtain 
$$
 \rho_{\Lambda(\lambda)}\Bigl(R_{H}^{[d],\im}(\lambda)\Bigr)=0.
$$
By using Lemma~\ref{uniquelemma} we have $R_{H}^{[d],\im}(\lambda)=0$. This produces a contradiction.  
The lemma is proved.
\end{proof}
Following~\cite{BDY21}, we define the {\it extended gradation} by assigning the following degrees:
\begin{align}
	&\overline{\deg}^{e} e_{i,j}=i-j,\quad \overline{\deg}^{e} \partial_{X}=1,\quad  \overline{\deg}^{e} \lambda =M,\quad i,j=1,\dots,M+1, \label{bt-extend-deg-1}\\
	&\overline{\deg}^{e} u_{l}=M-l,\quad l=-1,\dots,M-1.\label{bt-extend-deg-2}
\end{align}
Like in~\cite{BDY21} we have $\overline{\deg}^{e} \mathcal{L}(\lambda)=1$.
We are ready to prove Lemma~\ref{bt-MR-def}. 
\begin{proof}[Proof of Lemma~\ref{bt-MR-def}]
Let $U$ be the element defined in Lemma~\ref{fundmenlemma}. 
By using  Lemma \ref{kerlemma-ckp}, 
$
R_{a}(\lambda)=e^{U}\Lambda_{a}(\lambda)e^{-U},1\le a \le M
$ 
are solutions to equation~\eqref{defbtodreso}. 
Equation \eqref{reso} follows from \eqref{U-form}, and equation \eqref{carconchengbt} follows from the fact that
\begin{align}\label{leading}
	\tr (\Lambda_{a}(\lambda)\Lambda_{b}(\lambda))=\delta_{a+b,M}(a+b)\lambda+M\delta_{a+b,2M}\lambda^2.
\end{align}
This proves the existence of $R_{a}(\lambda)$. 
The uniqueness follows from Lemma \ref{kerlemma-ckp}. 
%The lemma is proved.		
\end{proof}

Recall that in Section~\ref{sec1}, 
we have defined a sequence of flows~\eqref{def-btflow} by using the basic matrix resolvents $R_{a}(\lambda)$, $1\le a\le M$. 
We are going to prove that~\eqref{def-btflow} coincide with~\eqref{laxeq-bt}.
\begin{lemma}\label{well-def-bt}
The flows \eqref{def-btflow} are well defined and coincide with the bigraded Toda hierarchy of $(M,1)$-type~\eqref{laxeq-bt}. 
\end{lemma}
\begin{proof}
Denote $\widehat{V}_{a,k}=-r_{a,k+1;M+1,1}e_{M+1,M+1}+\sum_{i=1}^{M-1}\sum_{j=i+1}^{M}\hat{v}_{a,k;i,j}e_{i,j},a=1,\dots,M,k\ge 0$.
By a direct calculation, we have
\begin{align}
		&	\mathcal{T}(V_{a,k})(\Lambda(\lambda)+V)-(\Lambda(\lambda)+V)V_{a,k} \nonumber\\
%		=&\mathcal{T}(R_{a,j+1})e_{1,M}-e_{1,M}R_{a,j+1}-\mathcal{T}(\widehat{V}_{a,j})(\Lambda(\lambda)+V)+(\Lambda(\lambda)+V)\widehat{V}_{a,j}\\
		=&\mathcal{T}(r_{a,k+1;1,1})e_{1,M}-\sum_{l=1}^{M+1}r_{a,k+1;M,l}e_{1,l}+\sum_{l=1}^{M-1}\mathcal{T}(\hat{v}_{a,k+1;1,l+1})e_{1,l}+\sum_{i=1}^{M-1}\sum_{l=i+1}^{M}u_{M-i}\hat{v}_{a,k+1;i,l}e_{1,l} \nonumber \\
		&-r_{a,k+1;M+1,1}u_{-1}e_{1,M+1}. \label{eq-bt}
\end{align}
From this expression, we can see that~\eqref{def-btflow} are well defined.
By using Lemma~\ref{bt-MR-def}, we have
\begin{align}
V_{1,0}(\lambda)=-\Lambda(\lambda)-V+ {\rm diag}\bigl(\mathcal{T}^{M-1}(\alpha_{M}), \dots, \alpha_{M}, \mathcal{T}^{-1}(\alpha_{M})\bigr),
%\begin{pmatrix}
%		\mathcal{T}^{M-1}(\alpha_{M})         &     &         &            \\
%		                &\ddots     &   &\\
%		    &      &  \alpha_{M}  &  \\
%		               &         &  &\mathcal{T}^{-1}(\alpha_{M})
%	\end{pmatrix},
\end{align}
where we recall the definition of $\Lambda(\lambda),V$ in~\eqref{def-lam-V} and  $\alpha_{M}:=(1+\mathcal{T}+\dots+\mathcal{T}^{M-1})^{-1}(u_{M-1})$. Then, by using~\eqref{def-ckpflow}, we have
\begin{align}
	\frac{\partial \mathcal{L}}{\partial t^{1}_{0}}=\epsilon^{-1}\sum_{i=1}^{M+1}\left((\mathcal{T}-1)(u_{i-3})+u_{i-2}(1-\mathcal{T}^{i-2})(\alpha_{M})\right)e_{1,M+2-i},\quad u_{-2}:=0.
\end{align}
This implies 
\begin{align}\label{flow-1}
\frac{\partial u_{k}}{\partial t^{1}_{0}}=\epsilon^{-1}\Bigl((\mathcal{T}-1)(u_{k-1})+u_{k}(1-\mathcal{T}^{k})(\alpha_{M})\Bigr),\quad k=-1,0,\dots,M-1.
\end{align}
This coincides with the $t^{1}_{0}$ flow of~\eqref{laxeq-cheng}. 
Then by using the method similar to~\cite{BDY16}, i.e., to match with the 
principal hierarchy of the corresponding Frobenius manifolds,  one can prove this lemma.  
\end{proof}
By using~\eqref{eq-bt} and~\eqref{def-btflow} we know that
the bigraded Toda hierarchy of $(M,1)$-type reads
\begin{align}
		&\epsilon\frac{\partial u_{i}}{\partial t_{k}^{a}}=-r_{a,k+1;M,M-i}+\mathcal{T}^{i+1}(r_{a,k+1;i+1,1})+\sum_{l=1}^{M-i-1}u_{M-l}\mathcal{T}^{i+1}(r_{a,k+1;i+l+1,1}),\\
		&\epsilon\frac{\partial u_{-1}}{\partial t_{k}^{a}}=u_{-1}(\mathcal{T}-1)(r_{a,k+1;M+1,1}),\label{flow-t1}
\end{align}
where $i=0,\dots,M,a=1,\dots,M,k\ge 0$.

\begin{lemma}
The following identities hold true
\begin{align}\label{eq1}
		&		\epsilon\frac{\partial R_{b}(\lambda)}{\partial t_{k}^{a}}=\left[V_{a,k}(\lambda),R_{b}(\lambda)\right],\quad k\ge 0, 1\le a,b\le M.
\end{align}
\end{lemma}
\begin{proof}
Using Lemma \ref{fundmenlemma}, we have
\begin{align}
		\epsilon\frac{\partial(\Lambda(\lambda)+H)}{\partial t_{k}^{a}}=\mathcal{T}\left(S_{a,k}\right)(\Lambda(\lambda)+H)-(\Lambda(\lambda)+H)S_{a,k},
\end{align}
where $S_{a,k}=e^{-\ad_{U}}(V_{a,k})-\frac{\partial U}{\partial t_{k}^{a}}$. 
Clearly, $S_{a,k}$ takes values in $\mathcal{A}[[\epsilon]]\otimes \mat(M+1,\mathbb{C}((\lambda^{-1})))$. 
Decompose 
\begin{align}
S_{a,k}= S_{a,k}^{\ke}+S_{a,k}^{\im},
\end{align}
where
$
S_{a,k}^{\ke} \in \mathcal{A}[[\epsilon]] \otimes \ke\,\tilde{\rho}_{\Lambda(\lambda)},S_{a,k}^{\im} \in \mathcal{A}[[\epsilon]] \otimes \im\,\tilde{\rho}_{\Lambda(\lambda)}.
$
Then we have
\begin{align}
		&\epsilon \frac{\p H}{\p t_{k}^{a} } =\mathcal{T}(S_{a,k}^{\ke})(\Lambda(\lambda)+H)-(\Lambda(\lambda)+H)S_{a,k}^{\ke} , \\  
		& 0 =\mathcal{T}(S_{a,k}^{\im})(\Lambda(\lambda)+H)-(\Lambda(\lambda)+H)S_{a,k}^{\im}  .\label{imeq}
\end{align}
Using the same argument as in the proof of Lemma~\ref{kerlemma}, 
we find that $S_{a,k}^{\im}$ must vanish. 
So, $S_{a,k}$ belongs to $\mathcal{A}[[\epsilon]] \otimes \ke\tilde{\rho}_{\Lambda(\lambda)}$. 
Then we have
\begin{align*}
		0=\Big[\epsilon\p_{t_{k}^{a}} - S_{a,k} , \Lambda_{b}(\lambda) \Big]=e^{-\ad_{U}}\Bigl[\epsilon \p_{t_{k}^{a}} - V_{a,k}(\lambda),R_{b}(\lambda) \Bigr].
\end{align*}
The lemma is proved.
\end{proof}
\begin{lemma}\label{bt-eq-zero}
We have the following identities
\begin{align}\label{bt-eq-zero-1}
		\epsilon	\nabla_{a}(\lambda)(R_{b}(\mu))=\frac{\left[R_{a}(\lambda),R_{b}(\mu)\right]}{\lambda-\mu}+\left[Q_{a}(\lambda),R_{b}(\mu)\right],\quad 1\le a,b\le M,
\end{align}
where 
\begin{align}
		Q_{a}(\lambda)=-\sum_{i=1}^{M-1}\sum_{j=i+1}^{M}\mathcal{T}^{M+1-j}(r_{a;M+1-j+i,1}(\lambda))e_{i,j}+r_{a;M+1,1}(\lambda)e_{M+1,M+1}-I\delta_{a,M}.
\end{align}
\end{lemma}
\begin{proof}	
By comparing the $\lambda$ coefficient of~\eqref{defbtodreso}, we have
	\begin{align}
	&r_{a,-1;i,i+k}=\mathcal{T}(r_{a,-1;i+1,i+1+k}),\quad r_{a,-1;M-k,M}=\mathcal{T}(r_{a,0;M+1-k,1}),\\
	& r_{a,-1;i,M+1}=0,\quad r_{M,0;M+1-k,1}=0,
	\end{align}
where $1\le i\le M+1-k, M+1-a\le k\le M-1,1\le a\le M$.
Then we have
\begin{align*}
		\sum_{k\ge 0}\frac{V_{a,k}(\mu)}{\lambda^{k+1}}
		%=&\sum_{k\ge 0}\frac{1}{\lambda^{k+1}}\sum_{i=-1}^{k}R_{a,i}\mu^{k-i}-\sum_{i=1}^{M+1}\sum_{j=1}^{M+1}\sum_{k\ge 0}\frac{\hat{v}_{a,k+1;i,j}}{\lambda^{k+1}}e_{i,j}\\
		=&\frac{R_{a}(\lambda)}{\lambda-\mu}-{\rm Coef}_{\lambda}(R_{a}(\lambda))\\
		&-\sum_{i=1}^{M-1}\sum_{j=i+1}^{M}\sum_{k\ge 0}\frac{\mathcal{T}^{M+1-j}(r_{a,k+1;M+1-j+i,1})}{\lambda^{k+1}}e_{i,j}+\sum_{k\ge 0}\frac{r_{a,k+1;M+1,1}}{\lambda^{k+1}}e_{M+1,M+1}\\
		=&\frac{R_{a}(\lambda)}{\lambda-\mu}-\sum_{i=1}^{M-1}\sum_{j=i+1}^{M}\mathcal{T}^{M+1-j}(r_{a;M+1-j+i,1}(\lambda))e_{i,j}+r_{a;M+1,1}(\lambda)e_{M+1,M+1}-I\delta_{a,M}.
\end{align*}
The lemma is proved by using~\eqref{eq1}.
\end{proof}
\begin{lemma}
We have the following identities
\begin{align}\label{eq2}
		\epsilon\frac{\partial V_{a,i}(\lambda)}{\partial t_{j}^{b}}-\epsilon\frac{\partial V_{b,j}(\lambda)}{\partial t_{i}^{a}}+\left[V_{a,i}(\lambda),V_{b,j}(\lambda)\right]=0,\quad 1\le a,b \le M, \quad i,j\ge 0.
\end{align}
\begin{proof}
Similar to the proof in Lemma~\ref{zero-ckp}, \eqref{eq2} are equivalent to  
\begin{align}\label{bt-q-ident}
	\epsilon\nabla_{b}(\mu)(Q_{a}(\lambda))-\epsilon\nabla_{a}(\lambda)(Q_{b}(\mu))+\left[Q_{a}(\lambda),Q_{b}(\mu)\right]=0.
\end{align}
Using~\eqref{bt-eq-zero-1}, we can obtain the following results by a direct calculation
\begin{align}\label{nabri1}
			\sum_{i=1}^{M+1}\epsilon\nabla_{a}(\lambda)(r_{b;i,1}(\mu))e_{i,1}=&\frac{1}{\lambda-\mu}\left(\sum_{i=1}^{M+1}\sum_{j=1}^{M+1}r_{a;i,j}(\lambda)r_{b;j,1}(\mu)e_{i,1}-\sum_{i=1}^{M+1}\sum_{j=1}^{M+1}r_{b;i,j}(\mu)r_{a;j,1}(\lambda)e_{i,1}\right)\nonumber\\
			&-\sum_{i=1}^{M-1}\sum_{j=i+1}^{M}\left[Q_{a}(\lambda)\right]_{i,j}r_{b;j,1}(\mu)e_{i,1}+r_{a;M+1,1}(\lambda)r_{b;M+1,1}(\mu)e_{M+1,1}.
\end{align}
Also we have
\begin{align}\label{nabri2}
			\sum_{i=1}^{M+1}\epsilon\nabla_{b}(\mu)(r_{a;i,1}(\lambda))e_{i,1}=&\frac{1}{\mu-\lambda}\left(\sum_{i=1}^{m}\sum_{j=1}^{m}r_{b;i,j}(\mu)r_{a;j,1}(\lambda)e_{i,1}-\sum_{i=1}^{M+1}\sum_{j=1}^{M+1}r_{a;i,j}(\lambda)r_{b;j,1}(\mu)e_{i,1}\right)\nonumber\\
			&-\sum_{i=1}^{M-1}\sum_{j=i+1}^{M}\left[Q_{b}(\mu)\right]_{i,j}r_{a;j,1}(\lambda)e_{i,1}+r_{b;M+1,1}(\mu)r_{a;M+1,1}(\lambda)e_{M+1,1}.
\end{align}
Substituting the above two equations~\eqref{nabri1}--\eqref{nabri2} into the left-hand side of~\eqref{bt-q-ident} and after a direct and tedious calculation, 
we can verify that~\eqref{bt-q-ident} indeed holds. 
We omit the details of the computation here.
The lemma is proved.
\end{proof}
\end{lemma}

Similarly as before, using~\eqref{eq2} one obtains the following lemma.
\begin{lemma}\label{comu-bt}
The flows~\eqref{def-btflow} pairwise commute.
\end{lemma}

\subsection{From matrix resolvents to tau-functions for the  bigraded Toda hierarchy of $(M,1)$-type}
Similarly to the proof in the Section~\ref{subsec2.2} (also cf.~\cite{BDY21,DY17}), 
we can show that the two-point correlation functions $\Omega_{a,i;b,j},1\le a,b\le M,i,j\ge 0$, for the bigraded Toda hierarchy, introduced in Section~\ref{sec1} (see equation~\eqref{def-bt-tau-str}), are well defined.
\begin{proof}[Proof of Lemma~\ref{tau-str-bt} ] By using~\eqref{bt-eq-zero-1}, the proof is similar to that in~\cite{BDY21,DY17}. We omit its details.
%The proof is similar to that in Lemma~\ref{tau-str-ckp} (also cf.~\cite{BDY21,DY17}), so we omit its details.
\end{proof}
\begin{proof}[Proof of Lemma~\ref{defbtodataufun}]
We first to prove the compatibility between \eqref{def-tau-1} and \eqref{def-tau-2}. On one hand, 
\begin{align*}
		&\sum_{i,j\ge 0}^{}\frac{1}{\lambda^{i+1}\mu^{j+1}}[\mathcal{T}(\Omega_{a,i;b,j})-\Omega_{a,i;b,j}]\\
		=&\frac{\tr \left(\mathcal{T}(R_{a}(\lambda))\mathcal{T}(R_{b}(\mu))-(\Lambda(\lambda)+V)^{-1}\mathcal{T}(R_{a}(\lambda))(\Lambda(\lambda)+V)(\Lambda(\mu)+V)^{-1}\mathcal{T}(R_{b}(\mu))(\Lambda(\mu)+V)\right)}{(\lambda-\mu)^{2}}\\
		=&\frac{\tr \left(\mathcal{T}(R_{a}(\lambda))\mathcal{T}(R_{b}(\mu))-(\Lambda(\mu)+V)(\Lambda(\lambda)+V)^{-1}\mathcal{T}(R_{a}(\lambda))(\Lambda(\lambda)+V)(\Lambda(\mu)+V)^{-1}\mathcal{T}(R_{b}(\mu))\right)}{(\lambda-\mu)^{2}}\\
		=&\frac{\tr \left(\mathcal{T}(R_{a}(\lambda))\mathcal{T}(R_{b}(\mu))-(I+(\mu-\lambda)e_{1,M+1})\mathcal{T}(R_{a}(\lambda))(I+(\lambda-\mu)e_{1,M+1})\mathcal{T}(R_{b}(\mu))\right)}{(\lambda-\mu)^{2}}\\
		=&\mathcal{T}\left(\frac{\sum_{j=1}^{M+1}r_{a;M+1,j}(\lambda)r_{b;j,1}(\mu)-\sum_{j=1}^{M+1}r_{b;M+1,j}(\mu)r_{a;j,1}(\lambda)}{\lambda-\mu}+r_{a;M+1,1}(\lambda)r_{b;M+1,1}(\mu)\right).
\end{align*}
On the other hand, it follows from \eqref{nabri1} that
\begin{align}\label{comptable}
		&\epsilon\nabla_{b}(\mu)\left(\mathcal{T}(r_{a;M+1,1}(\lambda))\right)\nonumber\\
		=&\mathcal{T}\left(\frac{\sum_{j=1}^{M+1}r_{a;M+1,j}(\lambda)r_{b;j,1}(\mu)-\sum_{j=1}^{M+1}r_{b;M+1,j}(\mu)r_{a;j,1}(\lambda)}{\lambda-\mu}+r_{a;M+1,1}(\lambda)r_{b;M+1,1}(\mu)\right).
\end{align}
Hence,
\begin{align}
		\sum_{i,j\ge 0}^{}\frac{1}{\lambda^{i+1}\mu^{j+1}}[\mathcal{T}(\Omega_{a,i;b,j})-\Omega_{a,i;b,j}]=\epsilon\nabla_{b}(\mu)(\mathcal{T}(r_{a;M+1,1}(\lambda))).
\end{align}
This proves the compatibility between \eqref{def-tau-1} and \eqref{def-tau-2}. 
By using~\eqref{comptable}, we have
\begin{align}
     	\epsilon\nabla_{b}(\mu)(\mathcal{T}(r_{a;M+1,1}(\lambda)))=\epsilon\nabla_{a}(\mu)(\mathcal{T}(r_{b;M+1,1}(\lambda)).
\end{align}
This proves the compatibility between all equations of~\eqref{def-tau-2}. 
Finally, we show the compatibility between \eqref{def-tau-3} and \eqref{def-tau-1},\eqref{def-tau-2}. Indeed,
\begin{align*}
		\sum_{i,j\ge 0}^{}\frac{1}{\lambda^{i+1}\mu^{j+1}}[\left(\mathcal{T}+\mathcal{T}^{-1}-2\right)(\Omega_{a,i;b,j})]=\left(\mathcal{T}-1\right)\epsilon\nabla_{b}(\mu)(r_{a;M+1,1}(\lambda)).
\end{align*}
Also,
\begin{align*}
		\epsilon^{2}\nabla_{b}(\mu)\nabla_{a}(\lambda)(\log u_{-1})=\epsilon\nabla_{b}(\mu)\left(\left(\mathcal{T}-1\right)(r_{a;M+1,1}(\lambda)-\delta_{a,M})\right).
\end{align*}
Hence,
\begin{align}
		\sum_{i,j\ge 0}^{}\frac{1}{\lambda^{i+1}\mu^{j+1}}[(\mathcal{T}+\mathcal{T}^{-1}-2)(\Omega_{a,i;b,j})]=	\epsilon^{2}\nabla_{b}(\mu)\nabla_{a}(\lambda)(\log u_{-1}).
\end{align}
This proves the compatibility between \eqref{def-tau-3} and \eqref{def-tau-1}.  
The compatibility between \eqref{def-tau-3} and \eqref{def-tau-2} is equivalent to \eqref{flow-t1}. 
The lemma is proved.
\end{proof}
Write $R_{a}({\lambda;x, {\bf t}};\epsilon)$ as the $R_{a}(\lambda)$ evaluated at an arbitrarily given solution~\eqref{solution-bt-sec1}.
The following corollary follows from Proposition~\ref{main-thm-bt}.
\begin{cor}
For any $k\ge 2$, we have
\begin{align}
		&\sum_{i_{1},\dots,i_{k}\ge 0}\epsilon^{k}\frac{\partial^{k} \log \tau(x,{\bf t};\epsilon)}{\partial t^{a_{1}}_{i_{1}}\cdots \partial t^{a_{k}}_{i_{k}}}\frac{1}{\prod_{j=1}^{k}\lambda_{j}^{i_{k}+1}} =-\sum_{\sigma\in S_{k}/C_{k}}\frac{\tr \bigl(R_{a_{\sigma(1)}}(\lambda_{\sigma(1)};x,{\bf t};\epsilon)\cdots R_{a_{\sigma(k)}}(\lambda_{\sigma(k)};x,{\bf t};\epsilon)\bigr)}{\prod_{i=1}^{k}(\lambda_{\sigma(i)}-\lambda_{\sigma(i+1)})} \nonumber\\
		&\qquad\qquad\qquad -\left(\frac{a_{1}\lambda_{1}+a_{2}\lambda_{2}}{(\lambda_{1}-\lambda_{2})^{2}}\delta_{a_{1}+a_{2},M}+M\frac{\lambda_{1} \lambda_{2}}{(\lambda_{1}-\lambda_{2})^{2}}\delta_{a_{1}+a_{2},2M}\right)\delta_{k,2}.
\end{align}
\end{cor}

\section{The relations between the constrained KP hierarchy and  the bigraded Toda hierarchy of $(M,1)$-type}\label{sec4}
In this section, based on the MR method, 
we study the relation between the constrained KP hierarchy and  the bigraded Toda hierarchy of $(M,1)$-type.

 For the case when $M=1$, the relation between the NLS hierarchy and the Toda lattice hierarchy is given in~\cite{CDZ04} (see also \cite{FY22}). 
 Following~\cite{FY22}, we look for the relation between 
 \begin{align}\label{def-a-bt}
 	\mathcal{A}(\lambda):=\epsilon\partial_{X}-V_{1,0}(\lambda)=\epsilon\partial_{X}+\Lambda(\lambda)+V-{\rm diag} \bigl(	\mathcal{T}^{M-1}(\alpha_{M}), \dots,  \alpha_{M} , \mathcal{T}^{-1}(\alpha_{M})\bigr),
% 	-\begin{pmatrix}
% 		\mathcal{T}^{M-1}(\alpha_{M})         &     &         &            \\
% 		&\ddots     &   &\\
% 		&      &  \alpha_{M}  &  \\
% 		&         &  &\mathcal{T}^{-1}(\alpha_{M})
% 	\end{pmatrix},
 \end{align}
and the Lax operator $\mathcal{L}_{\ckp}(\lambda)$,
where we recall that $\Lambda(\lambda),V$ are defined in~\eqref{def-lam-V} 
and that $\alpha_{M}=(1+\mathcal{T}+\dots+\mathcal{T}^{M-1})^{-1}(u_{M-1})$.

Start with the bigraded Toda hierarchy of $(M,1)$-type. 
Let 
$
u_{i}=u_{i}(x,{\bf t};\epsilon),i=-1,0,\dots,M-1,
$ 
be an arbitray solution to the  bigraded Toda hierarchy of $(M,1)$-type~\eqref{def-btflow}, 
and $\tau(x,{\bf t};\epsilon)$ the tau-function of this solution. We have the following lemma.
\begin{lemma}\label{key-pro}
There exists a unique $(M+1)\times (M+1)$ matrix $G$ which has the form
\begin{align}
		G=\begin{pmatrix}
			&                          &   & 0\\
			& \widehat{G}   &   &  \vdots\\
			&                          &   &  0\\
			0&  \cdots             &0 & -\frac{\tau(x-\epsilon,{\bf t};\epsilon)}{\tau(x,{\bf t};\epsilon)} 
		\end{pmatrix},
\end{align}
where $\widehat{G}$ is a $M\times M$ right-lower triangular matrix with the off-diagonal elements equal to $1$ and other elements in $\mathcal{A}[[\epsilon]]$, such that 
\begin{align}\label{keylaxrelation}
		G\circ \mathcal{A}(\lambda)\circ G^{-1}=\mathcal{L}_{\ckp}(\lambda).
\end{align}
\end{lemma}
\begin{proof}
Let
\begin{align}
	G_{1}=\begin{pmatrix}
		&      &    1 &    \\
		&  \begin{sideways}$\ddots$\end{sideways}  &     &    \\
		1 &      &   & \\
		&      &              &    -\frac{\tau(x-\epsilon,{\bf t};\epsilon)}{\tau(x,{\bf t};\epsilon)} 
	\end{pmatrix}.
\end{align}
We have
\begin{align}
	&G_{1}\circ\mathcal{A}(\lambda)\circ G_{1}^{-1} \nonumber\\
	=&\epsilon\partial_{X}+\left(\begin{array}{ccccccc|c}
		-\alpha_{M}        &   -1         & 0    & \cdots      &  0     &   0   &\cdots  &0\\
		0        & -\mathcal{T}(\alpha_{M})   & -1  &   \ddots     &       &     & &\vdots\\
		\vdots&   \ddots    &   \ddots       &   \ddots   &  \ddots     &    &   &\vdots\\
		\vdots        &      &             \ddots    &\ddots & \ddots     & 0&    &\vdots\\
		\vdots        &            &    &\ddots     &\ddots&-1  &0&0\\
		0        &     0       & 0  &   \cdots      & 0                        &-\mathcal{T}^{M-2}(\alpha_{M})&-1&0\\
		-\lambda+u_{0}  &   u_{1}    &  u_{3}      &   \cdots   &u_{M-3}&  u_{M-2}    & u_{M-1}-\mathcal{T}^{M-1}(r_{M-1})    &  -\frac{\tau(x+\epsilon,{\bf t};\epsilon)}{\tau(x,{\bf t};\epsilon)}\\
		\hline
		\frac{\tau(x-\epsilon,{\bf t};\epsilon)}{\tau(x,{\bf t};\epsilon)}       &   0         &   0 &\cdots  &  0     & 0     &0&0
	\end{array}\right),
\end{align}
Let $\widehat{M}^{A}(\lambda)$ be the $M\times M$ matrix operator given by $G_{1}\circ\mathcal{A}(\lambda)\circ G_{1}^{-1}$ by removing the $M+1$ row and $M+1$ column. 
Noticing $\tr \bigl(\widehat{M}^{A}(\lambda)-\epsilon\partial_{X}\bigr)=0$, 
there exists~\cite{BDY21, DS85} a unique strictly lower triangular matrix $N$ with entries in $\mathcal{A}[[\epsilon]]$
%$\widehat{G}_{2}$ with the diagonal elements equal to $1$ and other elements in $\mathcal{A}[[\epsilon]]$ 
such that 
$e^{N} \circ \widehat{M}^{A}(\lambda)\circ e^{-N}=\widehat{M}^{L}(\lambda)$,
%$
%	\widehat{G}_{2}\circ \widehat{M}^{A}(\lambda)\circ \widehat{G}_{2}^{-1}=\widehat{M}^{L}(\lambda),
%$
where $\widehat{M}^{L}(\lambda)$ has the form
\begin{align}
\widehat{M}^{L}(\lambda):=\epsilon\partial_{X}+\begin{pmatrix}
		        &   -1          &           &    & \\
		&           &   \ddots        &    &   \\
		&              &           & -1    &   \\
		        &             &       &&-1  \\
		-\lambda+v_{0}  &   v_{1}      &   \cdots   &  v_{M-1}    &  \\
	\end{pmatrix}.
\end{align}
Let $G_{2}=\begin{pmatrix}
	&                          &   & 0\\
	& e^{N}   &   &  \vdots\\
	&                          &   &  0\\
	0&  \cdots             &0 & 1 
\end{pmatrix}$
and let $G=G_{2}\circ G_{1}$,
the lemma is proved.
\end{proof}
\begin{remark}
Equation \eqref{keylaxrelation} implies that 
\begin{align}
q(x,{\bf t};\epsilon)=\frac{\tau(x+\epsilon,{\bf t};\epsilon)}{\tau(x,{\bf t};\epsilon)},\quad r(x,{\bf t};\epsilon)=\frac{\tau(x-\epsilon,{\bf t};\epsilon)}{\tau(x,{\bf t};\epsilon)}.
\end{align}
Then by using \eqref{def-tau-2}--\eqref{def-tau-3} we have
\begin{align}
	& \partial_{t_{k}^{a}}(\log q)=\epsilon^{-1}\mathcal{T}(r_{a,k+1;M+1,1}),\quad \partial_{t_{k}^{a}}(\log\,r)=-\epsilon^{-1}r_{a,k+1;M+1,1},   \label{tildeDj}	\\
	& qr=u_{-1},\quad \mathcal{T}^{-1}(q)r=1, \label{qrrelation}
\end{align}
where $1\le a\le M, k\ge 0$. In particular, when $a=1, k=0$, we have
\begin{align}
	\partial_{X}(\log q)=\epsilon^{-1}\alpha_{M},\quad\partial_{X}(\log r)=-\epsilon^{-1}\mathcal{T}^{-1}(\alpha_{M}).\label{0flow}
\end{align}
\end{remark}

\begin{lemma}
We have the following identities
\begin{align}\label{relamr}
		R_{a}^{\ckp}(\lambda)=G\left(R_{a}(\lambda)-\frac{M}{M+1}\lambda I\delta_{a,M}\right)G^{-1},\quad 1\le a\le M.
\end{align}
\end{lemma}
\begin{proof}
Denote $\widehat{R}_{a}(\lambda)=G\left(R_{a}(\lambda)-\frac{M}{M+1}\lambda I\delta_{a,M}\right)G^{-1}$. 
By using~\eqref{keylaxrelation}, we have
\begin{align}
		\left[\mathcal{L}_{\ckp}(\lambda),\widehat{R}_{a}(\lambda)\right]=G[\mathcal{A}(\lambda),R_{a}(\lambda)]G^{-1}.
\end{align}
By using \eqref{eq1}, \eqref{reso} and~\eqref{carconchengbt}, we know that 
\begin{align}
		&\left[\mathcal{L}_{\ckp}(\lambda),\widehat{R}_{a}(\lambda)\right]=0, \label{rstar1023-1}\\
		&\tr(\widehat{R}_{a}\widehat{R}_{b})=\delta_{a+b,M}M\lambda+\frac{M}{M+1}\lambda^{2}\delta_{a+b,2M} ,\\
		& \widehat{R}_{a}(\lambda)=\Lambda_{a}^{\ckp}(\lambda)+\cdots. \label{rstar1023-3}
\end{align}
By definition, $R_{a}^{\ckp}(\lambda)$ also satisfies~\eqref{rstar1023-1}--\eqref{rstar1023-3}. 
It is clear from the proof of Lemma~\ref{lemmarnls} that the solution to~\eqref{rstar1023-1}--\eqref{rstar1023-3} is unique. Therefore,  
$\widehat{R}_{a}(\lambda)=R_{a}^{\ckp}(\lambda)$. 
The lemma is proved.
\end{proof}
Let us now prove Theorem~\ref{maintheorem}.
\begin{proof}[Proof of Theorem~\ref{maintheorem}]
We first prove that the vector-valued function $(q,r,v_{0},\dots,v_{M-2})$ defined by~\eqref{keylaxrelation} 
is a solution to the constrained KP hierarchy \eqref{def-ckpflow} for any~$x$.
By using~\eqref{eq2},  we obtain
\begin{align}
	\frac{\partial \mathcal{A}(\lambda)}{\partial t_{k}^{a}}=\epsilon^{-1}[V_{a,k}(\lambda),\mathcal{A}(\lambda)],\quad a=1,\dots,M, ~ k\ge 0.\label{pequation}
\end{align}
Then by using \eqref{keylaxrelation} and \eqref{pequation}, for any $k\ge 0$, we obtain
\begin{align*}
	\frac{\partial \mathcal{L}_{\ckp}}{\partial t^{a}_{k}}=&\frac{\partial G}{\partial t^{a}_{k}}\mathcal{A}G^{-1}+G\frac{\partial \mathcal{A}}{\partial t^{a}_{k}}G^{-1}+G\mathcal{A}\frac{\partial G^{-1}}{\partial t^{a}_{k}}\\
	=&G\left[G^{-1}\frac{\partial G}{\partial t^{a}_{k}}+\epsilon^{-1}V_{a,k},\mathcal{A}\right]G^{-1}\\
	=&\left[\frac{\partial G}{\partial t^{a}_{k}}G^{-1}+\epsilon^{-1}GV_{a,k}G^{-1},\mathcal{L}_{\ckp}\right]\\
	=&\epsilon^{-1}[V_{a,k}^{\ckp},\mathcal{L}_{\ckp}].
\end{align*}
Here the last equality uses the uniqueness in Lemma~\ref{uni-refine}.

Substituting \eqref{relamr} into \eqref{def-ckp-tau-str}, and using \eqref{def-bt-tau-str}, we have
\begin{align}
	\sum_{i, j\ge 0} \frac{\Omega_{a,i;b,j}^{\ckp}}{\lambda^{i+1}\mu^{j+1}}=&\frac{\tr\left( R_{a}(\lambda)R_{b}(\mu)\right)}{(\lambda-\mu)^2}-\frac{a\lambda+b\mu}{(\lambda-\mu)^{2}}\delta_{a+b,M}-M\frac{\lambda \mu}{(\lambda-\mu)^{2}}\delta_{a+b,2M}=\sum_{i, j\ge 0} \frac{\Omega_{a,i;b,j}}{\lambda^{i+1}\mu^{j+1}}.\label{=1}		
\end{align}
Thus,
\begin{align}
	\Omega_{a,i;b,j}^{\ckp} %(x,{\bf t};\epsilon)
	=\Omega_{a,i;b,j} .%(x,{\bf t};\epsilon) . %=\epsilon^2\frac{\partial^2 \log\tau(x,{\bf t};\epsilon)}{\partial t_{i}^{a}\partial t_{j}^{b}},\quad i,j\ge 0,\label{=2}
\end{align}
The theorem is proved.
\end{proof}

Taking the $x$-derivative in~\eqref{keylaxrelation} we find 
\begin{align}
\label{xderiv}
	\frac{\p \mathcal{L}_{\ckp}(\lambda)}{\p x} = \frac{\p (G\circ \mathcal{A}(\lambda)\circ G^{-1})}{\p x}.
\end{align}
Let us define $v_{-1},v_{M-1}$ as in \eqref{v-1m-1}.
Using~\eqref{flow-1}, \eqref{v-1m-1}, \eqref{keylaxrelation} 
one can rewrite the $x$-flow~\eqref{xderiv} in terms of the dependent variables $v_{-1},\dots,v_{M-1}$ with $X$ being the space, which 
commutes with the flows of the constrained KP hierarchy written by using the variables $v_{-1},\dots,v_{M-1}$. 

\begin{remark}\label{rmk-inv}
We note that the transformation $(u_{-1},\dots,u_{M-1})\mapsto (v_{-1},\dots,v_{M-1})$ 
defined by~\eqref{keylaxrelation} (cf.~\eqref{v-1m-1}) is invertible, 
which can be seen by showing its triangular nature by a degree argument. 
\end{remark}

If we start with 
an arbitrary solution $(q({\bf t};\epsilon),r({\bf t};\epsilon),v_{0}({\bf t};\epsilon),\dots,v_{M-2}({\bf t};\epsilon))$ 
to the constrained KP hierarchy, 
by solving~\eqref{xderiv} (in the sense of the $x$-flow explained after~\eqref{xderiv}; note that the definition of 
$v_{-1}$ and $v_{M-1}$ from~\eqref{v-1m-1} only involves the $X$-derivative of $q({\bf t};\epsilon))$ 
and by considering Remark~\ref{rmk-inv}, we arrive at 
a solution $(u_{-1}(x,{\bf t};\epsilon),\dots,u_{M-1}(x,{\bf t};\epsilon))$ 
to the bigraded Toda hierarchy of $(M,1)$-type.
Define $q=q(x,{\bf t};\epsilon)$ by
\begin{align}
	&(1-\mathcal{T}^{-1})\log q=\log u_{-1},\\
	&  \delta_{a,M}+\sum_{i\ge 0}\frac{\epsilon}{\lambda^{i+1}}\frac{\partial \log q}{\partial t_{i}^{a}} =\bigl[R_{a}(\lambda;x+\epsilon,{\bf t};\epsilon)\bigr]_{M+1,1},
\end{align}
and define $r(x,{\bf t};\epsilon)=\frac{1}{q(x-\epsilon,{\bf t};\epsilon)}$,
then there exists a function $\tau_{\ckp}(x,{\bf t};\epsilon)$ satisfying 
\begin{align}
	&\epsilon^2\frac{\partial^2 \log\tau_{\ckp}(x,{\bf t};\epsilon)}{\partial t_{i}^{a}\partial t_{j}^{b}} = 
	\Omega_{a,i;b,j}^{\ckp}(x,{\bf t};\epsilon),\quad i,j\ge  0,\label{deftau1}\\
	&\epsilon (\mathcal{T}-1)\frac{\partial \log \tau_{\ckp}(x,{\bf t};\epsilon)}{\partial t_{k}^{a}}=\frac{1}{q}r_{a,k+1;1,M+1}^{\ckp}(x+\epsilon,{\bf t};\epsilon),\quad a=1,\dots,M,k\ge 0,\label{deftau2}\\
	&(\mathcal{T}+\mathcal{T}^{-1}-2)\log \tau_{\ckp}(x,{\bf t};\epsilon)=\log (q(x,{\bf t};\epsilon)r(x,{\bf t};\epsilon)).\label{deftau3}
\end{align}
The compatibility between \eqref{deftau1}, \eqref{deftau2} and \eqref{deftau3} can be proved by using 
\begin{align}\label{rx+}
	R_{a}^{\ckp}(x+\epsilon;{\bf t};\lambda;\epsilon)\mathcal{T}(G)(\Lambda(\lambda)+V)G^{-1}-\mathcal{T}(G)(\Lambda(\lambda)+V)G^{-1}R_{a}^{\ckp}(x,{\bf t};\lambda;\epsilon)=0,
\end{align}
which is similar to that in the proof of Lemma \ref{defbtodataufun}, so we omit its details.
Then by using \eqref{=1}, we have
\begin{align}
		&\epsilon^2\frac{\partial^2 \log\tau_{\ckp}(x,{\bf t};\epsilon)}{\partial t_{i}^{a}\partial t_{j}^{b}}=\Omega_{i,j}(x,{\bf t};\epsilon),\quad i,j\ge  0.\label{identitytodatau1}
\end{align}
This together with \eqref{deftau2},~\eqref{deftau3} and~\eqref{relamr} implies that
$\tau_{\ckp}(x,{\bf t};\epsilon)$ is the tau-function of the solution
\begin{align}\label{solu-bt}
(u_{-1}(x,{\bf t};\epsilon),\dots,u_{M-1}(x,{\bf t};\epsilon))
\end{align}	
to the bigraded Toda hierarchy.

\begin{example} 
The case $M=2$.
We have
\begin{align}
		\mathcal{L}_{\ckp}=\epsilon\partial_{X}+\Lambda^{\ckp}(\lambda)+V_{\ckp},\quad \Lambda=\begin{pmatrix}
			0&-1&0\\
			-\lambda &0&0\\
			0&0&0
		\end{pmatrix},\quad V_{\ckp}=\begin{pmatrix}
			0&0&0\\
			v_{0}&0&-q\\
			r&0&0
		\end{pmatrix}.
\end{align}
The basic matrix resolvents of $\mathcal{L}_{\ckp}$ are
\begin{align}
		R_{1}^{\ckp}(\lambda)=&\begin{pmatrix}
			0&0&0\\
			1&0&0\\
			0&0&0
		\end{pmatrix}\lambda+\begin{pmatrix}
			0&1&0\\
			-\frac{v_{0}}{2}&0&q\\
			-r&0&0
		\end{pmatrix}+\mathcal{O}(\lambda^{-1}), \\
		R_{2}^{\ckp}(\lambda)=&\begin{pmatrix}
			\frac{1}{3}&0&0\\
			0 &\frac{1}{3}&0\\
			0&0&-\frac{2}{3}
		\end{pmatrix}\lambda+\begin{pmatrix}
			0&0&q\\
			0&0&\epsilon q_{X}\\
			\epsilon r_{X}&-r&0
		\end{pmatrix}+\mathcal{O}(\lambda^{-1}).
\end{align}
The first few flows of the constrained KP hierarchy~\eqref{def-ckpflow} are given by
\begin{align}
	&\frac{\partial v_{0}}{\partial t_{0}^{1}}=v_{0,X},\quad \frac{\partial q}{\partial t_{0}^{1}}=q_{X}\quad \frac{\partial r}{\partial t_{0}^{1}}=r_{X},\label{ckp-3-flows-1}\\
	&\frac{\partial v_{0}}{\partial t_{0}^{2}}=2(qr)_{X},\quad  \frac{\partial q}{\partial t_{0}^{2}}=\epsilon q_{2X}+q v_{0},\quad \frac{\partial r}{\partial t_{0}^{2}}=-(\epsilon r_{2X}+\epsilon^{-1}r v_{0}), \label{y-o-hier}\\
	&\frac{\partial v_{0}}{\partial t_{1}^{1}}=\frac{1}{4} \left(6\epsilon r q_{2X}-6\epsilon q r_{2X}+\epsilon^{2}v_{0,3X}+6 v_{0} v_{0,X}\right),\\
	& \frac{\partial q}{\partial t_{1}^{1}}=\frac{1}{4} \left(4\epsilon^{2} q_{3X}+6 v_{0} q_{X}+6\epsilon^{-1} q^2 r+3 q v_{0,X}\right),\\
	& \frac{\partial r}{\partial t_{1}^{1}}=\frac{1}{4} \left(-6\epsilon^{-1} qr^2+4\epsilon^{2} r_{3X}+6 v_{0} r_{X}+3 r v_{0,X}\right).
\end{align}
Here \eqref{y-o-hier} are the equations given by  Yajima and Oikawa~\cite{YO76}.
From the definition~\eqref{def-ckp-tau-str} the first few terms of the tau-structure for the constrained KP hierarchy are
\begin{align*}
&\Omega_{1,0;1,0}^{\ckp}=\frac{v_{0}}{2},\quad \Omega_{1,0;2,0}^{\ckp}=qr,\quad \Omega_{2,0;2,0}^{\ckp}=\epsilon(r q_{X}-q r_{X}),\\
&\Omega_{1,1;1,0}^{\ckp}=\frac{1}{8} \left(3 v_{0}^2+6\epsilon (r q_{X}-q r_{X})+\epsilon^{2} v_{0,2X}\right).
%&\Omega_{1,1;2,0}^{\ckp}=\frac{3}{2}q r v_{0}+\epsilon^{2}(r q_{2X}-q_{X} r_{X}+q r_{2X}),\\
\end{align*}

The matrix-valued Lax operator of the bigraded Toda hierarchy of $(2,1)$-type reads
\begin{align}
	\mathcal{L}=\mathcal{T}+\Lambda(\lambda)+V,\quad 	\Lambda(\lambda)=\begin{pmatrix}
		0&-\lambda &0\\
		-1&0&0\\
		0&-1&0
	\end{pmatrix},\quad V=\begin{pmatrix}
		u_{1}&u_{0}&u_{-1}\\
		0&0&0\\
		0&0&0
	\end{pmatrix}.
\end{align}
The basic matrix resolvents of $\mathcal{L}$ have the form
\begin{align}
	R_{1}(\lambda)=&\begin{pmatrix}
		0&1&0\\
		0&0&0\\
		0&0&0
	\end{pmatrix}\lambda+
	\begin{pmatrix}
		-\alpha_{2}&-(1+\mathcal{T}^{-1})^{-1}(\mathcal{T}^{-1}(u_{0})+\alpha_{2}^{2})&-u_{-1}\\
		1&\alpha_{2}&0\\
		0&1&0
	\end{pmatrix}+\mathcal{O}(\lambda^{-1}),\\
	R_{2}(\lambda)=&\begin{pmatrix}
		1&0&0\\
		0&1&0\\
		0&0&0
	\end{pmatrix}\lambda+\begin{pmatrix}
		0&0&0\\
		0&0&-u_{-1}\\
		1&\mathcal{T}^{-1}(u_{1})&0
	\end{pmatrix}+\mathcal{O}(\lambda^{-1}),
\end{align}
where $\alpha_{2}=(1+\mathcal{T})^{-1}(u_{1})$.
The first few flows of the bigraded Toda hierarchy~\eqref{def-btflow} of $(2,1)$-type are
\begin{align}
	&\epsilon\frac{\partial u_{1}}{\partial{t_{0}^{1}}}=(\mathcal{T}-1)(u_{0})+u_{1}(1-\mathcal{T})\left(\alpha_{2}\right),\qquad \epsilon\frac{\partial u_{1}}{\partial{t_{0}^{2}}}=(\mathcal{T}^{2}-1)(u_{-1}),\\
	&\epsilon\frac{\partial u_{0}}{\partial{t_{0}^{1}}}=(\mathcal{T}-1)(u_{-1}),\qquad \epsilon\frac{\partial u_{0}}{\partial{t_{0}^{2}}}=\mathcal{T}(u_{-1})u_{1}-u_{-1}\mathcal{T}^{-1}(u_{1}),\\
	&\epsilon\frac{\partial u_{-1}}{\partial{t_{0}^{1}}}=u_{-1}(1-\mathcal{T}^{-1})(\alpha_{2}),\qquad \epsilon\frac{\partial u_{-1}}{\partial{t_{0}^{2}}}=u_{-1}(1-\mathcal{T}^{-1})(u_{0}).
\end{align}
Using the definition~\eqref{def-bt-tau-str} we have
\begin{align*}
&\Omega_{1,0;1,0}=(1+\mathcal{T})^{-1}(u_{0}-\alpha_{2}^{2}),\quad \Omega_{1,0;2,0}=u_{-1},\quad \Omega_{2,0;2,0}=u_{-1}\mathcal{T}^{-1}(u_{1}),\\
&\Omega_{1,1;2,0}=u_{-1}(\mathcal{T}^{-1}(u_{0})+(1+\mathcal{T}^{-1})^{-1}(u_{0}-\alpha_{2}^{2})+\alpha_{2}\mathcal{T}^{-1}(u_{1})).
\end{align*}
The matrix operator $\mathcal{A}$ defined in~\eqref{def-a-bt} is given by
\begin{align}
	\mathcal{A}=\epsilon\partial_{X}-V_{1,0}=\epsilon\partial_{X}-\begin{pmatrix}
		-u_{1}+\mathcal{T}(\alpha_{2})&\lambda-u_{0}&-u_{-1}\\
		1&\alpha_{2}&0\\
		0&1&\mathcal{T}^{-1}(\alpha_{2})
	\end{pmatrix}.
\end{align}
The unique matrix $G$ given in Lemma~\ref{key-pro} reads
\begin{align}
	G=\begin{pmatrix}
		0&1&0\\
		1&\alpha_{2}&0\\
		0&0&-\frac{\mathcal{T}^{-1}(\tau)}{\tau}
	\end{pmatrix}.
\end{align}
We have $G\mathcal{A}G^{-1}=\mathcal{L}_{\ckp}$, where $(q,r,v_{0})$ and $(u_{-1},u_{0},u_{1})$ 
are related by~\eqref{qr-def-603}--\eqref{v0-def-603}.
The statements in Theorem~\ref{maintheorem} can then be verified straightforwardly.
\end{example}

\begin{appendices}
\section{Proofs of Lemma~\ref{ker-im-ckp}  and  Lemma~\ref{ker-im-bt} }\label{appen-A}
\begin{proof}[Proof of Lemma~\ref{ker-im-ckp} ]
 Firstly, we will prove $e_{l}^{\ckp}\in \im\ad_{\Lambda^{\ckp}(\lambda)}, 1\le l\le M^{2}+M$.
Define
\begin{align*}
			&E_{k(M-1)+i}^{\ckp}=\left\{\begin{aligned}
				&	- \sum_{q=1}^{i}e_{q+1,q+k}\quad  & i=1,\dots ,M-k-1\\
				&	e_{1,k}+\lambda\sum_{l=M-k}^{M-2}e_{l+2,l-(M-k)+1}\quad       & i=M-k,\dots, M-1
			\end{aligned}  \right. ,\quad k=1,\dots, M-1,\\
			&E_{M(M-1)+1}^{\ckp}=\frac{1}{\lambda}e_{M+1,M},\\
			& E_{M(M-1)+i+1}^{\ckp}=e_{M+1,i},\quad E_{(M+1)(M-1)+i+1}^{\ckp}=-e_{i+1,M+1}, \quad i=1,\dots, M-1.
\end{align*}
It is easy to see that $E_{l}^{\ckp}\in sl_{M+1}(\mathbb{C})((\lambda^{-1})),1\le l\le M^{2}+M$. By a direct calculation, we have 
\begin{align}
			\ad_{\Lambda^{\ckp}(\lambda)}(E_{l}^{\ckp})=e_{l}^{\ckp},\quad l=1,\dots,M^{2}+M.
\end{align}	
This implies that $e_{l}^{\ckp}\in \im\ad_{\Lambda^{\ckp}(\lambda)}$. It is also easy to see that $\Lambda_{a}^{\ckp}\in \ke\ad_{\Lambda^{\ckp}(\lambda)},1\le a\le M$. 
Secondly, we will prove the linearly independence of $e_{l}^{\ckp},\Lambda_{a}^{\ckp},1\le l\le M^{2}+M,1\le a\le M$. Suppose that there exist $(M+1)^{2}-1$ constants $a_{k,i},b_{a},c_{a},d_{a},d_{M+1},0\le k\le M-1,1\le i\le M-1,1\le a\le M$ such that
\begin{align}\label{basis-ckp}
			&\sum_{k=0}^{M-1}\sum_{i=1}^{M-1}a_{k,i}e_{k(M-1)+i}^{\ckp}+\sum_{a=1}^{M}b_{a}e_{M(M-1)+a}^{\ckp}+\sum_{a=1}^{M}c_{a}e_{(M+1)(M-1)+a+1}^{\ckp}+\sum_{a=1}^{M}d_{a}\Lambda_{a}^{\ckp}=0.
\end{align}
It follows that
\begin{align}
	        & b_{a}=c_{a}=0,\quad a=1,\dots, M,\\
			& \sum_{i=1}^{M-1}a_{0,i}+\frac{M}{M+1}\lambda d_{M}=0,\\
			&-a_{0,i}+\frac{M}{M+1}\lambda d_{M}=0,\quad i=1,\dots, M-1,\\
			& \sum_{i=1}^{M-1} a_{k,i}+d_{k}=0,\quad k=1,\dots, M-1,\\
			&a_{k,i}+d_{k}=0,\quad k,i=1,\dots,M-1.
\end{align}
These imply that $a_{k,i}=d_{a}=0, 0\le k,i\le M-1,1\le a\le M$. We have proved the linear independence.
Thirdly, we will prove that $e_{l}^{\ckp},1\le l\le M^{2}+M$ is a basis of $\im\ad_{\Lambda^{\ckp}(\lambda)}$. Indeed, for any $A\in sl_{M+1}(\mathbb{C})((\lambda^{-1}))$, write $A=\sum_{i=1}^{M+1}\sum_{j=1}^{M+1}A_{i,j}e_{i,j}$. Then we have
\begin{align}
			\ad_{\Lambda^{\ckp}(\lambda)}(A)=&\sum_{i=1}^{M-1}\sum_{j=2}^{M}(A_{i,j-1}-A_{i+1,j})e_{i,j}+\sum_{i=1}^{M-1}(\lambda A_{i,M}-A_{i+1,1})e_{i,1}+\sum_{j=2}^{M}(A_{M,j-1}-\lambda A_{1,j})e_{M,j} \nonumber\\
			&+\lambda (A_{M,M}-A_{1,1})e_{M,1}-\lambda A_{1,M+1}e_{M,M+1}+\lambda A_{M+1,M}e_{M+1,1}-\sum_{i=1}^{M-1}A_{i+1,M+1}e_{i,M+1}  \nonumber\\
			&+\sum_{j=2}^{M}A_{M+1,j-1}e_{M+1,j}.\label{im-ckp}
\end{align} 
Denote $\sum_{i=1}^{M+1}\sum_{j=1}^{M+1}B_{i,j}e_{i,j}:=\ad_{\Lambda^{\ckp}(\lambda)}(A)$. Then we have
\begin{align}
			\sum_{j=1}^{M-i}B_{j+i,j}=-\lambda \sum_{j=1}^{i}B_{j,M-i+j},\quad i=0,\dots,M-1.
\end{align}
These imply $\dim_{\mathbb{C}((\lambda^{-1}))}\im\ad_{\Lambda^{\ckp}(\lambda)}\le (M+1)^{2}-1-M=M^{2}+M$. Since $e_{l}^{\ckp}\in \im\ad_{\Lambda^{\ckp}(\lambda)},1\le l\le M^{2}+M$ are linearly independent. Then we have that $\dim_{\mathbb{C}((\lambda^{-1}))}\im\ad_{\Lambda^{\ckp}(\lambda)}=M^{2}+M$ and $e_{l}^{\ckp}$ is a basis of $ \im\ad_{\Lambda^{\ckp}(\lambda)}$. Finally, we will prove that $\Lambda_{a}^{\ckp},1\le a\le M$ is a basis of $\ke \ad_{\Lambda^{\ckp}(\lambda)}$. Indeed, for any $A\in \ke \ad_{\Lambda^{\ckp}(\lambda)}$ satisfies $\ad_{\Lambda^{\ckp}(\lambda)}(A)=0$, we have, by using~\eqref{im-ckp}, that
\begin{align}
			&A_{M+1,l}=A_{l,M+1}=0,\quad l=1,\dots,M,\\
			&A_{1,1+i}=A_{2,2+i}=\cdots=A_{M-i,M}=\frac{1}{\lambda}A_{M-i+1,1}=\frac{1}{\lambda}A_{M-i+2,2}=\cdots=\frac{1}{\lambda}A_{M,i}, \quad i=0,\dots,M-1.
\end{align}
%where $ i=0,\dots,M-1$. 
Thus $\dim_{\mathbb{C}((\lambda^{-1}))} \ke \ad_{\Lambda^{\ckp}(\lambda)}=(M+1)^{2}-1-(M+1)M=M$. Since $\Lambda_{a}^{\ckp}, 1\le a\le M$, are linearly independent, we have that $\Lambda_{a}^{\ckp}, 1\le a\le M$, is a basis of $ \ke\ad_{\Lambda^{\ckp}(\lambda)}$. The lemma is proved.
\end{proof}

\begin{proof}[Proof of Lemma~\ref{ker-im-bt}]
We first to prove that $e_{l}\in \im\tilde{\rho}_{\Lambda(\lambda)}, 1\le l\le M^{2}+M$. Define
\begin{align*}
		&E_{k(M-1)+i}=\left\{\begin{aligned}
			&	\lambda \sum_{q=1}^{i}e_{q+k,q+k+1}\quad  & i=1,\dots ,M-k-1\\
			&	-\sum_{j=0}^{M-i}e_{M+1-j,k+2-j}\quad       & i=M-k,\dots, M-1
		\end{aligned}  \right. , \quad k=1,\dots, M-1, \\
		& E_{M(M-1)+i}=e_{i,1}-\lambda e_{i,M+1},\quad E_{(M+1)(M-1)+i+1}=e_{M+1,i+1}\quad i=1,\dots,M,\\
		& E_{(M+1)(M-1)+1}=-\sum_{l=1}^{M}e_{l,l+1}.
\end{align*}
By a direct calculation, we have 
\begin{align}
		\tilde{\rho}_{\Lambda(\lambda)}(E_{l})=e_{l},\quad l=1,\dots,M^{2}+M.
\end{align}
This implies that $e_{l}\in \im\tilde{\rho}_{\Lambda(\lambda)}$. 
It is easy to see that $\Lambda_{a}(\lambda),I\in \ke\tilde{\rho}_{\Lambda(\lambda)},1\le a\le M$. 
Secondly, we will prove the linear independence of $e_{l},\Lambda_{a}(\lambda), I,1\le l\le M^{2}+M,1\le a\le M$.  
Suppose there exist $(M+1)^{2}-1$ constants $a_{k,i},b_{a},c_{a},d_{a},d_{M+1},0\le k\le M-1,1\le i\le M-1,1\le a\le M$ such that
\begin{align}\label{basis}
		&\sum_{k=0}^{M-1}\sum_{i=1}^{M-1}a_{k,i}e_{k(M-1)+i}+\sum_{a=1}^{M}b_{a}e_{M(M-1)+a}+\sum_{a=1}^{M}c_{a}e_{(M+1)(M-1)+a+1}+\sum_{a=1}^{M}d_{a}\Lambda_{a}+d_{M+1}I=0.
\end{align}
It follows that
\begin{align}
	    &b_{a}=d_{M+1}=0,\quad a=1,\dots,M,\\
		& c_{k+1}+d_{M-k}=0,\\
		& \sum_{i=1}^{M-1}a_{k,i}+d_{M-k}-\delta_{k,0}Md_{M}=0,\\
		&-a_{k,i}+d_{M-k}-\delta_{k,0}Md_{M}=0,
\end{align}
where $k=0,\dots,M,  i=1,\dots,M-1$.
These imply that $a_{k,i}=c_{a}=d_{a}=0, 0\le k\le M-1,1\le i\le M-1,1\le a\le M$.
We have proved the linear independence. 
Thirdly, we will prove that $e_{l},1\le l\le M^{2}+M$ is a basis of $\im\tilde{\rho}_{\Lambda(\lambda)}$. 
Indeed, for any element $A\in  \mat(M+1,\mathbb{C}((\lambda^{-1})))$, write $A=\sum_{i=1}^{M+1}\sum_{j=1}^{M+1}A_{i,j}e_{i,j}$. 
Then we have
\begin{align}\label{im-bt}
		\tilde{\rho}_{\Lambda(\lambda)}(A)=&\sum_{i=2}^{M+1}\sum_{j=1}^{M}(A_{i,j+1}-A_{i-1,j})e_{i,j}+\sum_{j=1}^{M}(A_{1,j+1}-\lambda A_{M,j})e_{1,j}-\lambda A_{M,M+1}e_{1,M+1}\nonumber\\
		&+\lambda\sum_{i=1}^{M+1}A_{i,1}e_{i,M}-\sum_{i=2}^{M+1}A_{i-1,M+1}e_{i,M+1}.
\end{align}
Denote $\sum_{i=1}^{M+1}\sum_{j=1}^{M+1}B_{i,j}e_{i,j}:=\ad_{\Lambda}(A)$. Then we have
\begin{align}
		&B_{1,M+1}+\lambda B_{M+1,M+1}=0,\quad \sum_{k=1}^{M+1}B_{k,k}=0,\\
		&\lambda\sum_{j=1}^{M-i}B_{j+i,j}+\sum_{j=1}^{i+1}B_{j,M-i+j}=0,\quad i=1,\dots,M-1.
\end{align}
These imply $\dim_{\mathbb{C}((\lambda^{-1}))}\im\tilde{\rho}_{\Lambda(\lambda)}\le M^{2}+M$. Since $e_{l}\in \im\tilde{\rho}_{\Lambda(\lambda)},1\le l\le M^{2}+M$ are linearly independent, we have that $\dim_{\mathbb{C}((\lambda^{-1}))}\im\tilde{\rho}_{\Lambda(\lambda)}=M^{2}+M$ and $e_{l}$ is a basis of $ \im\ad_{\Lambda}$. Finally, we will prove that $I,\Lambda_{a}(\lambda),1\le a\le M$ is a basis of $\ke \tilde{\rho}_{\Lambda(\lambda)}$. Indeed, for any element $A\in \ke \tilde{\rho}_{\Lambda(\lambda)}$, we have $\tilde{\rho}_{\Lambda(\lambda)}(A)=0$. By using~\eqref{im-bt} we have 
\begin{align}
		&A_{1,1}=A_{2,2}=\cdots=A_{M,M}=\frac{1}{\lambda}A_{M+1,1}+A_{M+1,M+1},\quad A_{p,M+1}=0,\quad p=1,\dots,M,\\
		&A_{i+1,1}=A_{i+2,2}=\cdots=A_{M+1,M+1-i}=\frac{1}{\lambda}A_{1,M+1-i}=\frac{1}{\lambda}A_{2,M-i+2}=\cdots=\frac{1}{\lambda}A_{i,M},\quad i=1,\dots,M-1.
\end{align}
Thus $\dim_{\mathbb{C}((\lambda^{-1}))} \ke \tilde{\rho}_{\Lambda(\lambda)}=(M+1)^{2}-M(M+1)=M+1$. Since $I,\Lambda_{a}(\lambda), 1\le a\le M$ are linearly independent, we have that $I,\Lambda_{a}(\lambda), 1\le a\le M$, form a basis of $ \ke\tilde{\rho}_{\Lambda(\lambda)}$. The lemma is proved.
\end{proof}

\end{appendices}

\medskip
\medskip
\medskip

\medskip

\noindent Ang Fu

\noindent fuang@mail.ustc.edu.cn

\medskip

\noindent Di Yang

\noindent diyang@ustc.edu.cn

\medskip

\noindent Dafeng Zuo

\noindent dfzuo@ustc.edu.cn

\medskip

\noindent School of Mathematical Sciences, University of Science and Technology of China

\noindent Hefei 230026, P.R.~China

\end{document}